\documentclass[structabstract]{aa}  
\usepackage{geometry}
\usepackage{graphicx}
\usepackage{color}
\usepackage{amssymb}
\usepackage{natbib}
\usepackage{lscape}

\newcommand{\mc}[3]{\multicolumn{#1}{#2}{#3}}

\usepackage{txfonts}
\usepackage[dvipdfm]{hyperref}
\hypersetup{breaklinks=true, colorlinks=true, citecolor=blue, linkcolor=black}
\begin{document}
   \title{The F-GAMMA program: Multi-frequency study of Active Galactic Nuclei
in the {\it Fermi} era}

   \subtitle{Program description and the first 2.5 years of monitoring}

   \author{L.~Fuhrmann\inst{1}, E.~Angelakis\inst{1}, J.~A.~Zensus\inst{1}, I. Nestoras\inst{1},
     N.~Marchili\inst{2}, V.~Pavlidou\inst{1,3,4}, V.~Karamanavis\inst{1},
     H.~Ungerechts\inst{5}, T.~P.~Krichbaum\inst{1}, S.~Larsson\inst{6,7}, 
     S.~S.~Lee\inst{10}, W.~Max-Moerbeck\inst{1},
     I.~Myserlis\inst{1}, T.~J.~Pearson\inst{4}, A.~C.~S.~Readhead\inst{4}, J.~L.~Richards\inst{4,8},
     A.~Sievers\inst{5}, B.~W.~Sohn\inst{10,11}}

   \institute{
     Max-Planck-Institut f\"ur Radioastronomie, Auf dem H\"ugel 69, D-53121 Bonn, Germany
     \and
     IAPS-INAF, Via Fosso del Cavaliere 100, 00133, Roma, Italy
     \and
     Department of Physics/ Institute for Plasma Physics, University of Crete, and FORTH-IESL, 71003,
     Heraklion, Greece
     \and
     California Institute of Technology, Pasadena, CA 91125, USA
     \and
     Instituto de Radioastronom\'ia Milim\'etrica, Avenida Divina Pastora 7, Local 20, 18012 Granada, Spain
     \and
     KTH Royal Institute of Technology, Department of Physics, AlbaNova, SE-10691 Stockholm, Sweden
     \and
     Oskar Klein Centre, Department of Astronomy, AlbaNova, SE-10691 Stockholm, Sweden
     \and
     Department of Physics, Purdue University, 525 Northwestern Ave, West Lafayette, IN 47907
     \and
     Korea Astronomy and Space Science Institute, Daedeokdae-ro 76, Yuseong, Daejeon 305-348, Republic of Korea
     \and
     University of Science and Technology, 133 Gwahangno, Yuseong-gu, Daejeon, 305-333, Republic of Korea
}
   
   \date{Received ---, ---; accepted ---, ---}

 
  \abstract
  {To fully exploit the scientific potential of the {\it Fermi} mission for AGN physics, we initiated the
    F-GAMMA program. Between 2007 and 2015 the F-GAMMA was the prime provider of complementary multi-frequency
    monitoring in the radio regime.}
  {We quantify the radio variability of $\gamma$-ray blazars. We investigate its dependence on source class and
    examine whether the radio variability is related to the $\gamma$-ray loudness. Finally, we assess the
    validity of a putative correlation between the two bands.}
  {The F-GAMMA monitored monthly a sample of about 60 sources at up to twelve radio frequencies between 2.64
    and 228.39~GHz. We perform a time series analysis on the first 2.5-year dataset to obtain variability
    parameters. A maximum likelihood analysis is used to assess the significance of a correlation between
    radio and $\gamma$-ray fluxes.}
  {We present light curves and spectra (coherent within ten days) obtained with the Effelsberg 100-m and IRAM
    30-m telescopes. All sources are variable across all frequency bands with amplitudes increasing with
    frequency up to rest frame frequencies of around 60--80~GHz as expected by shock-in-jet models. Compared
    to FSRQs, BL\,Lacs show systematically lower variability amplitudes, brightness temperatures and Doppler
    factors at lower frequencies, while the difference vanishes towards higher ones. The time scales appear
    similar for the two classes. The distribution of spectral indices appears flatter or more inverted at
    higher frequencies for BL\,Lacs. Evolving synchrotron self-absorbed components can naturally account for
    the observed spectral variability. We find that the {\it Fermi}-detected sources show larger variability
    amplitudes as well as brightness temperatures and Doppler factors, than non-detected ones. Flux densities
    at 86.2 and 142.3~GHz correlate with 1~GeV fluxes at a significance level better than $3\sigma$, implying
    that $\gamma$ rays are produced very close to the mm-band emission region.}
{}

   \authorrunning{Fuhrmann et al.}
   \titlerunning{The F-GAMMA program: Program description and the first 2.5 years of monitoring}
   \maketitle
%

\section{Introduction}
\label{sec:introduction}



Powerful jets, sometimes extending outwards to several Mpc from a 
bright nucleus, are the most striking feature of radio-loud active galactic 
nuclei (AGN). It is theorised that the power sustaining these
systems is extracted through
the infall of galactic material onto a supermassive black hole via an accretion 
flow. Material is then 
channeled to
jets
which transport angular momentum and energy 
away from
the active nucleus in the intergalactic space
\citep[][]{1974MNRAS.169..395B,1974MNRAS.166..305H}.  Unification theories attribute the phenomenological
variety of AGN types to the combination of their intrinsic energy output and
orientation of their jets relative to our line-of-sight
\citep[][]{1978Natur.276..768R,1979ApJ...232...34B,1980PhyS...21..662R, 1995PASP..107..803U}.

Blazars, viewed at angles not larger than $10^{\circ}$ to $20^{\circ}$, form the sub-class of radio-loud AGN
showing the most extreme phenomenology. This typically involves strong broadband variability, high degree of
optical polarisation, apparent superluminal motions
\citep[e.g.][]{1966ApJ...144..843D,1971ApJ...170..207C,1972ApJ...173L.147S} and a unique broadband,
double-humped spectral energy distribution (SED) \citep[][]{1999APh....11..159U}. Moreover, blazars have long
been established as a group of bright and highly variable $\gamma$-ray sources
\citep[e.g.][]{1992ApJ...385L...1H}.  While different processes may occur in different objects, the
high-energy blazar emission is believed to be produced by the inverse Compton (IC) mechanism acting on seed
photons inside the jet (synchrotron self-Compton, SSC) or external Compton (EC) acting on photons from the
broad-line region (BLR) or the accretion disk \citep[leptonic models,
e.g.][]{1992A&A...256L..27D,2007Ap&SS.309...95B}. Alternatively, proton induced cascades and their products
have been invoked to account for it, in the case of hadronic jets \citep[e.g.][]{1993A&A...269...67M}.


Despite all efforts, several key questions still remain unanswered. For example:
(i) which are the dominant, broadband emission processes, (ii) which mechanisms drive the violent, broadband
variability of blazars and (iii) what is the typical duty cycle of their activity?
A number of competing models attempt to explain their observed properties in terms of e.g. relativistic
shock-in-jet models \citep[e.g.][]{1985ApJ...298..114M,1992A&A...254...71V,2011MmSAI..82..104T} or colliding
relativistic plasma shells \citep[e.g.][]{2001MNRAS.325.1559S,2004A&A...421..877G}.  Quasi-periodicities seen
in the long-term variability curves on time scales of months to years, may indicate systematic changes in the
beam orientation \citep[lighthouse effect:][]{1992A&A...255...59C}, possibly related to binary black hole
systems, MHD instabilities in the accretion disks and/or helical/precessing jets
\citep[e.g.][]{1980Natur.287..307B,1993A&A...279...83C,1999A&A...347...30V}.  Finally, the location of
$\gamma$-ray emission is still intensely debated \citep[cf.][]{1995ApJ...441...79B,
  1995A&A...297L..13V,2001ApJ...556..738J,2014ApJ...780...87M}.


Analysis and theoretical modelling of (quasi-simultaneous) spectral variability, over spectral ranges as broad
as possible (radio to GeV/TeV energies), allows the detailed study of different emission mechanisms and
comparison with different competing theories.
Hence, variability studies furnish important clues about the size, structure, physics and dynamics of the
emitting region making AGN/blazar monitoring programs extremely important in providing the necessary
constraints for understanding the origin of energy production.




Examples of past and ongoing 
long-term
radio monitoring programs with variability studies are the University of Michigan Radio Observatory (UMRAO)
program \citep[4.8--14.5~GHz; e.g.][]{1970ApJ...161....1A,1985ApJS...59..513A}, the monitoring at the
Medicina/Noto
32-m telescopes \citep[5, 8, 22~GHz;][]{2007A&A...464..175B}, and the RATAN-600 monitoring
\citep[1--22~GHz;][]{2002PASA...19...83K}; all at lower radio frequencies.  At intermediate frequencies, the
Mets\"ahovi Radio Observatory
program \citep[22 \& 37~GHz; e.g.][]{1987A&AS...70..409S,1988A&A...203....1V}, while at high
frequencies the 
IRAM 30-m monitoring program \citep[90, 150, 230~GHz;
e.g.][]{1988A&AS...75..317S,1992A&AS...96..441S,1998ASPC..144..149U}.
Nevertheless, the lack of continuous observations at all wave bands and the historical lack of sufficient
$\gamma$-ray data prevented past efforts from studying in detail the broadband jet emission.

The launch of the {\it Fermi} Gamma-ray Space Telescope ({\it Fermi}) in June 2008, with its high-cadence
``all-sky monitor'' capabilities, has introduced a new era in the field of AGN astrophysics providing a
remarkable opportunity for attacking the crucial questions outlined above. The Large Area Telescope
\citep[LAT;][]{2009ApJ...697.1071A} on-board {\it Fermi} constitutes a large leap
compared to its predecessor, the Energetic Gamma-ray Experiment Telescope (EGRET). Since 2008, {\it Fermi} has
gathered spectacular $\gamma$-ray spectra and light curves resolved at a variety of time scales for
about 1500 AGN
\citep[e.g.][]{2010ApJ...722..520A,2011ApJ...743..171A,2015ApJ...810...14A}.


The {\it Fermi}-GST AGN Multi-frequency Monitoring Alliance (F-GAMMA) represents an effort, highly-coordinated
with {\it Fermi} and other observatories, for the monitoring of selected AGN. Here, we present F-GAMMA and
report on the results obtained for the initial sample, during the first 2.5 years of observations (January
2007 to June 2009).  
The paper is structured as follows. In Sect.~\ref{sec:gamma-project}, we introduce the program, discuss the
sample selection, describe the participating observatories and outline the data
reduction. Sections~\ref{sct:Variabilityanalysis},~\ref{fgamma_fermi} and \ref{flux_flux_corr} present the
variability analysis, the connection between radio variability and $\gamma$-ray loudness, and the radio and
$\gamma$-ray flux--flux correlation, respectively. We summarise our results and conclude in
Sect.~\ref{sect:Summary}.

\section{The F-GAMMA program}
\label{sec:gamma-project}
The F-GAMMA program \citep[][]{2007AIPC..921..249F,2010arXiv1006.5610A,2014MNRAS.441.1899F,
  2015A&A...575A..55A} aimed at providing a systematic monthly monitoring of the radio emission of about 60
$\gamma$-ray blazars over the frequency range from 2.64 to 345 GHz. The motivation was the acquisition of
uniformly sampled light curves meant to:
\begin{itemize}
\item complement the {\it Fermi} light curves, and other, ideally simultaneous, datasets for the construction
  of coherent SEDs \citep[e.g.][]{2012A&A...541A.160G},
\item be used for studying the evolutionary tracks of spectral components as well as variability studies
  \cite[e.g. shock models][]{1992A&A...254...71V,2015A&A...580A..94F},
\item be used for cross-band correlations and time series analyses \citep[e.g.][]{2014MNRAS.441.1899F,2016A&A...590A..48K} and
\item correlations with source structural evolution studies \citep{2016A&A...586A..60K}.
\end{itemize}
Below we explain the sample selection and then the observations and the data reduction for each facility
separately.


\subsection{Sample selection}
\label{subsec:sample}

\begin{table*}
  \caption{The initial F-GAMMA sample monitored at EB and PV between January
2007 and June 2009. The last column indicates whether a source is included other programs}   
\label{tab:catalog}  
\centering                    
{\scriptsize 
\begin{tabular}{@{\hskip .1cm}l@{\hskip .1cm}l@{\hskip .15cm}l@{\hskip .15cm}l@{\hskip .15cm}l@{\hskip .15cm}l@{\hskip .15cm}l@{\hskip .1cm}l@{\hskip .15cm}l@{\hskip .15cm}l@{\hskip .15cm}l@{\hskip .15cm}l@{\hskip .1cm}} 
\hline
\hline                 
ID           & Catalog         & Class\tablefootmark{a}  & R.A.        & DEC         &Other programs\tablefootmark{b} &ID           & Catalog         & Class\tablefootmark{a}  & R.A.        & DEC         &Other programs\tablefootmark{b}\\       
             & Name            &        & (J2000)     & (J2000)     &                                &             & Name            &         & (J2000)    & (J2000)     &                               \\       
\hline\\                          
J0006$-$0623 &                 & FSRQ    & 00:06:13.9 & $-$06:23:35 & C M Pl W Po                               &J1159$+$2914 &                 & FSRQ    & 11:59:31.8 & $+$29:14:44 & B C F M Pl W Po F1 F2      \\        
J0102$+$5824 &                 & FSRQ    & 01:02:45.8 & $+$58:24:11 & M Po F2                 		        &J1221$+$2813 & WCom  ON231     & BL Lac  & 12:21:31.7 & $+$28:13:59 & B C E F M Pl Po F1 F2      \\        
J0217$+$0144 & PKS0215$+$015   & FSRQ    & 02:17:49.0 & $+$01:44:50 & C E F M Pl Po F1 F2             	        &J1224$+$2122 & 4C21.35 	& FSRQ    & 12:24:54.5 & $+$21:22:46 & B E M Po F2		   \\	    
J0222$+$4302 & 3C66A           & BL Lac  & 02:22:39.6 & $+$43:02:08 & B F M Pl W Po F1 F2             	        &J1229$+$0203 & 3C273		& FSRQ    & 12:29:06.7 & $+$02:03:09 & B C E F G M Pl W Po F1 F2 \\      
J0237$+$2848 & 4C28.07         & FSRQ    & 02:37:52.4 & $+$28:48:09 & C E F M Pl W Po F1 F2           	        &J1230$+$1223 & M87		& RG	  & 12:30:49.4 & $+$12:23:28 & M W Po F2		\\	    
J0238$+$1636 & AO0235$+$16     & BL Lac  & 02:38:38.9 & $+$16:36:59 & B C E F G M Pl W Po F1 F2       	        &J1256$-$0547 & 3C279		& FSRQ    & 12:56:11.2 & $-$05:47:22 & B C E F G M Pl W Po F1 F2	\\  
J0241$-$0815 & NGC1052         & Sy 2    & 02:41:04.8 & $-$08:15:21 & C M Pl Po                             	&J1310$+$3220 & OP313           & blazar  & 13:10:28.7 & $+$32:20:44 & B C F M Pl W Po F1 F2     \\   
J0303$+$4716 & 4C47.08         & BL Lac  & 03:03:35.2 & $+$47:16:16 & M Po F2                      		&J1408$-$0752 & PKS1406$-$076   & FSRQ    & 14:08:56.5 & $-$07:52:27 & B E M Pl W Po F2               \\	    
J0319$+$4130 & 3C84            & blazar  & 03:19:48.2 & $+$41:30:42 & C F M W Po F1 F2               	        &J1428$+$4240 & H1426$+$428     & BL Lac  & 14:28:32.7 & $+$42:40:21 & Po F2                     	\\    
J0319$+$1845 & 1E0317.0$+$1835 & BL Lac  & 03:19:51.8 & $+$18:45:34 & C Po F2                     		&J1504$+$1029 & PKS1502+106     & FSRQ    & 15:04:25.0 & $+$10:29:39 & C F M Pl W F1 F2             	\\    
J0336$+$3218 & OE355           & FSRQ    & 03:36:30.1 & $+$32:18:29 & E M Pl Po F2                 		&J1512$-$0905 & PKS1510$-$08    & FSRQ    & 15:12:50.5 & $-$09:05:60 & B E F G M Pl W Po F1 F2       \\     
J0339$-$0146 & PKS0336$-$01    & FSRQ    & 03:39:30.9 & $-$01:46:36 & B C M Pl W Po F2               		&J1522$+$3144 &                 & FSRQ    & 15:22:10.0 & $+$31:44:14 & C F M F1 F2                 	\\    
J0359$+$5057 & NRAO150         & FSRQ    & 03:59:29.7 & $+$50:57:50 & M Po                      		&J1540$+$8155 & 1ES1544$+$820   & BL Lac  & 15:40:16.0 & $+$81:55:06 & Po                     		\\    
J0418$+$3801 & 3C111           & Sy 1    & 04:18:21.3 & $+$38:01:36 & B E G M Po F2               	        &J1613$+$3412 & OS319		& FSRQ    & 16:13:41.1 & $+$34:12:48 & B C E M Pl W Po F2		\\    
J0423$-$0120 & PKS0420$-$01    & FSRQ    & 04:23:15.8 & $-$01:20:33 & B C G M Pl W Po F2             	        &J1635$+$3808 & 4C38.41 	& FSRQ    & 16:35:15.5 & $+$38:08:05 & B C E F G M Pl W Po F1 F2	\\     
J0433$+$0521 & 3C120           & blazar  & 04:33:11.1 & $+$05:21:16 & C E M Pl W Po                 	        &J1642$+$3948 & 3C345		& FSRQ    & 16:42:58.8 & $+$39:48:37 & B C G M Pl W Po F2		\\     
J0507$+$6737 & 1ES0502$+$675   & BL Lac  & 05:07:56.3 & $+$67:37:24 & F Po F1 F2                 		&J1653$+$3945 & Mkn501          & BL Lac  & 16:53:52.2 & $+$39:45:37 & C F M Pl W Po F1 F2           \\     
J0530$+$1331 & PKS0528$+$134   & FSRQ    & 05:30:56.4 & $+$13:31:55 & B C E F G M Pl Po F1 F2         	        &J1733$-$1304 & NRAO530 	& FSRQ    & 17:33:02.7 & $-$13:04:50 & B C E M Pl Po F2		\\    
J0721$+$7120 & S50716$+$71     & BL Lac  & 07:21:53.4 & $+$71:20:36 & B C E F G M Pl W Po F1 F2       	        &J1800$+$7828 & S51803$+$78	& BL Lac  & 18:00:45.7 & $+$78:28:04 & C F M Pl W Po F1 F2		\\     
J0738$+$1742 & PKS0735$+$17    & BL Lac  & 07:38:07.4 & $+$17:42:19 & B C E F M Pl W Po F1 F2         	        &J1806$+$6949 & 3C371		& BL Lac  & 18:06:50.7 & $+$69:49:28 & C M Pl W Po F2			\\    
J0750$+$1231 &                 & FSRQ    & 07:50:52.0 & $+$12:31:05 & C M Pl W Po F2                 		&J1824$+$5651 & 4C56.27         & BL Lac  & 18:24:07.1 & $+$56:51:01 & C M Pl W Po F2                 	\\    
J0818$+$4222 & TXS0814$+$425   & BL Lac  & 08:18:16.0 & $+$42:22:45 & C F M Pl Po F1 F2             		&J1959$+$6508 & 1ES1959$+$650   & BL Lac  & 19:59:59.9 & $+$65:08:55 & C F M Po F1 F2                 	\\    
J0830$+$2410 & OJ248           & FSRQ    & 08:30:52.1 & $+$24:10:60 & B C E G M Pl W Po F2             		&J2158$-$1501 & PKS2155$-$152   & FSRQ    & 21:58:06.3 & $-$15:01:09 & C M Pl W Po F2                 	\\    
J0841$+$7053 & S50836$+$71     & FSRQ    & 08:41:24.4 & $+$70:53:42 & B G M W Po F2               		&J2158$-$3013 & PKS2155$-$304   & BL Lac  & 21:58:52.0 & $-$30:13:32 & C F Pl Po F1 F2             	\\    
J0854$+$2006 & OJ287           & BL Lac  & 08:54:48.9 & $+$20:06:31 & B C E F G M Pl W Po F1 F2       	        &J2202$+$4216 & BLLac           & BL Lac  & 22:02:43.3 & $+$42:16:40 & B E F G M Pl W Po F1 F2       \\     
J0958$+$6533 & S40954$+$65     & BL Lac  & 09:58:47.2 & $+$65:33:55 & B C E M Pl Po F2             		&J2203$+$3145 & 4C31.63         & FSRQ    & 22:03:15.0 & $+$31:45:38 & C M Pl W Po                 	\\    
J1041$+$0610 & PKS1038$+$064   & FSRQ    & 10:41:17.2 & $+$06:10:17 & C M Pl W Po                 		&J2225$-$0457 & 3C446           & FSRQ    & 22:25:47.3 & $-$04:57:01 & B G M Pl W Po F2               	\\    
J1104$+$3812 & Mkn421          & BL Lac  & 11:04:27.3 & $+$38:12:32 & C E F M Pl Po F1 F2             	        &J2232$+$1143 & OY150  CTA102	& FSRQ    & 22:32:36.4 & $+$11:43:51 & B E F G M Pl W Po F1 F2	\\  
J1130$-$1449 & PKS1127$-$14    & FSRQ    & 11:30:07.1 & $-$14:49:27 & B F M Pl W Po F1 F2             	        &J2253$+$1608 & 3C454.3 	& FSRQ    & 22:53:57.7 & $+$16:08:54 & B C E F G M Pl W Po F1 F2	\\  
J1128$+$5925 &                 & FSRQ    & 11:28:13.0 & $+$59:25:15 & C                      			&J2347$+$5142 & 1ES2344$+$514   & BL Lac  & 23:47:04.8 & $+$51:42:18 & M Po                     	\\	    
J1136$+$7009 & Mkn180          & BL Lac  & 11:36:26.4 & $+$70:09:27 & C Pl Po F2                 		&J2348$-$1631 & PKS2345$-$16    & FSRQ    & 23:48:02.6 & $-$16:31:12 & C M Pl W Po                      \\ \\   
\hline
\end{tabular}
  \tablefoot{
    \tablefoottext{a}{Source class: FSRQs=flat-spectrum radio quasar, BL Lac=BL\,Lac object, Sy1/2=Seyfert
      type 1/2, RG= radio galaxy, blazar=unidentified blazar type.}
    \tablefoottext{b}{B=Boston 43~GHz VLBI program, C=CGRABS sample, E=EGRET detected (3EG catalog), G=GMVA 86~GHz VLBI 
    program, M=MOJAVE program, Pl=Planck detected, W=WMAP detected, F1={\it Fermi} detected (LBAS), 
    F2={\it Fermi} detected (1LAC), Po=POLAMI monitoring program.}
}
}
\end{table*}

 The selection of the F-GAMMA source sample was determined by our goal to understand the physical processes in
$\gamma$-ray-loud blazars; in particular their broadband variability and spectral evolution during periods of
energetically violent outbursts. By definition, the program was designed to take advantage of the continuous
{\it Fermi} monitoring of the entire sky. Since the F-GAMMA observations started more than a year before the
{\it Fermi} launch, the monitored sample was subjected to a major update (in mid 2009) once the first {\it
  Fermi} lists were released to include only sources monitored by the satellite.


Initially, the F-GAMMA sample included 62 sources selected to satisfy several criteria. The most important of
which, were:
\begin{enumerate}
\item Be previously detected in $\gamma$ rays by EGRET and be included in the ``high priority AGN/blazar
  list'' released by the {\it Fermi}/LAT AGN working group, which would tag them as potential {\it Fermi}
  $\gamma$-ray candidates.
\item Display flat radio spectra, an identifying characteristic of the blazars behaviour.
\item Comply with certain observational constrains: (a) be at relatively high declination
  ($\delta \ge -30^\circ$), and (b) give high average brightness to allow uninterruptedly reliable and high
  quality data flow.
\item Show frequent activity -- in as many energy bands as possible -- to allow cross-band and variability
  studies. 

\end{enumerate}

In Table~\ref{tab:catalog}, we list the sources in the initial sample that were observed with the Effelsberg
and IRAM telescopes between January 2007 and June 2009, which is the period covered by the present work. The
Effelsberg and IRAM observations were done monthly and in a highly synchronised manner. A sub-set of 24 of
these sources and an additional sample of about 20 southern $\gamma$-ray AGNs were observed also with the APEX
telescope \citep[see][]{2012arXiv1206.3799L}.

An important consideration during the sample selection has been the overlap with other campaigns and
particularly VLBI monitoring programs. Most of the 62 sources in our sample (95\%) are included in the IRAM
polarisation monitoring \cite[POLAMI; e.g.][]{2014A&A...566A..59A}. A major fraction (89\%) were observed by
MOJAVE \citep{2009AJ....137.3718L}, and almost half (47\%) are in the Boston 43~GHz program
\citep{2005AJ....130.1418J}. One third of our sources (27\%) are in the GMVA monitoring and one source (namely
PKS\,2155-304) is part of the southern TANAMI VLBI program \citep{2010A&A...519A..45O}. The 3$^{rd}$ EGRET
catalog \citep[][]{1999ApJS..123...79H} includes 40\% of the initial sample sources while the majority of
sources (69\%) are in CGRaBS. WMAP point source catalog \citep[][]{2003ApJS..148...97B} includes 61\% of our
sample and 74\% are in the Planck Early Release Compact Source Catalogue
\citep[ERCSC,][]{2011A&A...536A...7P}. The last column Table~\ref{tab:catalog} summarises the overlap with all
these programs.

On the basis of the classical AGN classification scheme: flat-spectrum radio quasars (FSRQs), BL Lacertae
objects (BL Lacs) and radio galaxies \citep[e.g. ][]{1995PASP..107..803U}, the initial F-GAMMA sample consists
of 32 FSRQs (52\%), 23 BL Lacs (37\%), 3 radio galaxies (5\%) and 3 unclassified blazars (5\%).

In June 2009, soon after the release of the first {\it Fermi} source lists \citep[LBAS and
1LAC;][]{2009ApJ...700..597A,2010ApJ...715..429A}, the F-GAMMA sample was seriously revised to include
exclusively {\it Fermi} monitored sources. The revised list will be presented in Nestoras et al. (revised) and
Angelakis et al. (in preparation).

\subsection{Observations and data reduction}

The main facilities employed for the F-GAMMA program were the Effelsberg 100-m 
(hereafter EB), IRAM 30-m (PV) and APEX
12-m telescopes. The observations were highly coordinated between the EB and PV. As it was already discussed
beyond those, several other facilities and teams have occasionally participated in campaigns in collaboration
wit the F-GAMMA program or have provided complementary studies. Here we summarise the involved facilities, the
associated data acquisition and reduction.
\begin{table*}
  \caption{{\bf Top:} The participating facilities of the F-GAMMA program. {\bf Bottom:} telescopes of 
  main collaborations and complementary projects.}     
  \label{tab:facilities}  
  \centering     
  \small                
  \begin{tabular}{l l l l l l} 
    \hline\hline             
    Facility &Location &Band  &Diameter &Frequency         &Notes \\    
             &         &      &(m)      &(GHz)             &        \\    
\hline\\
Effelsberg 100-m      & Effelsberg, DE        &110--7~mm &100 &2.64, 4.85, 8.35, 10.45, 14.6 &intensity \& polarisation \\   
                      &                       &          &    &23.05, 32, 43                 &intensity \\ 
IRAM 30-m             & Sierra Nevada, ES     &3--1~mm   &30  &86.2, 142.3, 228.9            &intensity\\   
APEX 12-m             & Atacama Plateau, CL   &0.85~mm   &12  &345                           &quasi-regular since 2007\\\\

\hline\\
OVRO 40-m             & Owens Valley, CA      &20~mm     &40  &15.0                          &2--3 times per week\\  
KVN                   & Korea                 &13, 7~mm  &21  &21.7, 42.4                    &monthly since 2010 \\    
IRAM 30-m             & Sierra Nevada, ES     &3, 1~mm   &30  &86.2, 228.9                   &monthly, polarisation \\
ROBOPOL               & Crete, GR             &optical   &1.3 &$R$-band                      &intensity \& polarisation \\\\
    \hline                                  
  \end{tabular}
\end{table*}

\subsubsection{The Effelsberg 100-m telescope}
\label{subsec:EFF}
The most important facility for the F-GAMMA program has been the Effelsberg 
100-m telescope owing to: the broad frequency coverage, the large number of 
available receivers in that range and the high
measurement precision. 

The observations started in January 2007 and ceased in January 2015. Here however we focus only on the first
2.5 years over which the initial F-GAMMA sample was monitored. The program was being scheduled monthly in 30
to 40-hour-long sessions. The observations were conducted with the eight secondary focus receivers listed in
Table~\ref{tab:facilities} covering the range from 2.64 to 43~GHz in total power and polarisation (whenever
available). Their detailed characteristics are listed in Table~4 of \cite{2015A&A...575A..55A}.



The measurements were done with ``cross-scans'' i.e. by measuring the telescope response as it progressively
slews over the source position in azimuth and elevation direction. This technique allows the correction of
small pointing offsets, the detection of possible spatial extension of a source or cases of confusion from
field sources. The systems at 4.85, 10.45 and 32~GHz are equipped with multiple feeds which were used for
subtracting tropospheric effects (``beam switch''). The time needed for obtaining a whole spectrum was of the
order of 35--40 minutes, guaranteeing measurements free of source variability.

The data reduction and the post-measurement data processing is described in Section~3 of
\cite{2015A&A...575A..55A}. For the absolute calibration we used the reference sources 3C\,48, 3C\,161,
3C\,286, 3C\,295 and NGC\,7027 \citep[][]{1977A&A....61...99B,1994A&A...284..331O,2008ApJ...681.1296Z}. The
assumed flux density values for each frequency are listed in Table~3 of \cite{2015A&A...575A..55A}. The
overall measurement uncertainties are of the order of ${\le} 1\%$ and ${\le} 5\%$ at lower and higher
frequencies, respectively.
 



\subsubsection{The IRAM 30-m telescope}
\label{fgamma_PV}
The IRAM 30-m telescope at Pico Veleta covered the important short-mm bands.  The observations started in June
2007 and ceased in May 2014 using the ``B'' and ``C'' SIS heterodyne receivers at 86.2 (500~MHz bandwidth),
142.3 and 228.9~GHz both with 1~GHz bandwidth (Table~\ref{tab:facilities}). The receivers were operated in a
single linear polarisation mode but simultaneously at the 3 observing frequencies.

To minimise the influence of blazar variability in the combined spectra, the EB and PV observations were
synchronised typically within a few days up to about one week. The F-GAMMA observations were combined with the
general flux monitoring conducted by IRAM \citep[e.g.][]{1998ASPC..144..149U}. The data presented here come
from both programs.

Observations were performed with calibrated cross-scans in the azimuth and elevation directions. For each
target the cross-scans were preceded by a calibration scan to obtain instantaneous opacity information and
convert the counts to the antenna temperature scale $T_{\mathrm{A}}^{*}$ corrected for atmospheric attenuation
as described by \cite{1989A&AS...79..217M}. After a necessary quality flagging, the sub-scans in each scanning
direction were averaged and fitted with Gaussian curves. Each fitted amplitude was then corrected for pointing
offsets and elevation-dependent losses.  The absolute calibration was done with reference to frequently
observed of primary calibrators: Mars, Uranus and secondary ones: W3(OH), K3-50A, NGC\,7027.  The overall
measurement uncertainties are 5--10\%. 





\subsection{Other facilities}
In order to accommodate the needs for broader frequency coverage, dense monitoring of large unbiased samples,
and synchronous monitoring of structural evolution a number of collaborating facilities were coordinated with
the F-GAMMA program. For completeness their are briefly described below. 

\noindent
{\it The APEX sub-mm telescope:}
In 2008, we incorporated the usage of the Large Apex Bolometer Camera \citep[LABOCA;][]{2008SPIE.7020E...2S}
of the APEX (Atacama Pathfinder Experiment) 12-m telescope to obtain light curves at 345~GHz for a subset of
25 F-GAMMA sources and 14 southern hemisphere $\gamma$-ray AGN.
In the current we do not present APEX data. First results are however discussed in
\citet[][]{2012arXiv1206.3799L}.

\noindent
{\it The OVRO 40-m program:}
In middle 2007 a program complementary to F-GAMMA was commenced at the OVRO 40-m telescope. It is dedicated to
the dense monitoring of a large sample of blazars at 15~GHz. The initial sample included 1158 CGRaBS sources
\citep[][]{2008ApJS..175...97H} with declination $\ge-20^{\circ}$. Later the sample was enriched with {\it
  Fermi} to include around 1800 sources. The cadence is around a few days. The details of the OVRO program are
discussed by \cite{2011ApJS..194...29R,2014MNRAS.438.3058R}.

\noindent
{\it Korean VLBI Network program:}
The AGN group of the Korean VLBI Network (KVN) has been using the KVN antennas since 2010 for the monthly
monitoring $\gamma$-ray blazars simultaneously at 22 and 42~GHz \citep[][]{2011PASP..123.1398L}. Details of
the observing method can be found in \citet[][]{2013arXiv1311.3038P}.

\noindent
{\it Optical monitoring:}
In collaboration with the California Institute of Technology (Caltech), the University of Crete, the
Inter-University Centre for Astronomy and Astrophysics (IUCAA) and the Torun Center for Astronomy at the
Nicolaus Copernicus University - we have initiated, funded and constructed the RoboPol optical polarimeter
\citep[][]{2014MNRAS.442.1706K} mounted on the 1.3-m Skinakas optical telescope (Greece). Since 2013 RoboPol
has been monitoring the optical linear polarisation of a sample of 60 blazars including 47 F-GAMMA
sources. First polarisation results of RoboPol are presented in
\citet[][]{2014MNRAS.442.1693P,2015MNRAS.453.1669B,2016MNRAS.457.2252B}.

\section{Variability analysis}
\label{sct:Variabilityanalysis}
Here, we study the initial F-GAMMA sample. We focus on the first 2.5 years of EB 
observations (31 sessions from January 2007 to June 2009) and the first 2 years 
monitoring with PV (21 sessions from June 2007 to June 2009).

\subsection{Light curves and variability characteristics}
\begin{figure*}
  \centering
  \includegraphics[angle=0,width=0.9\textwidth]{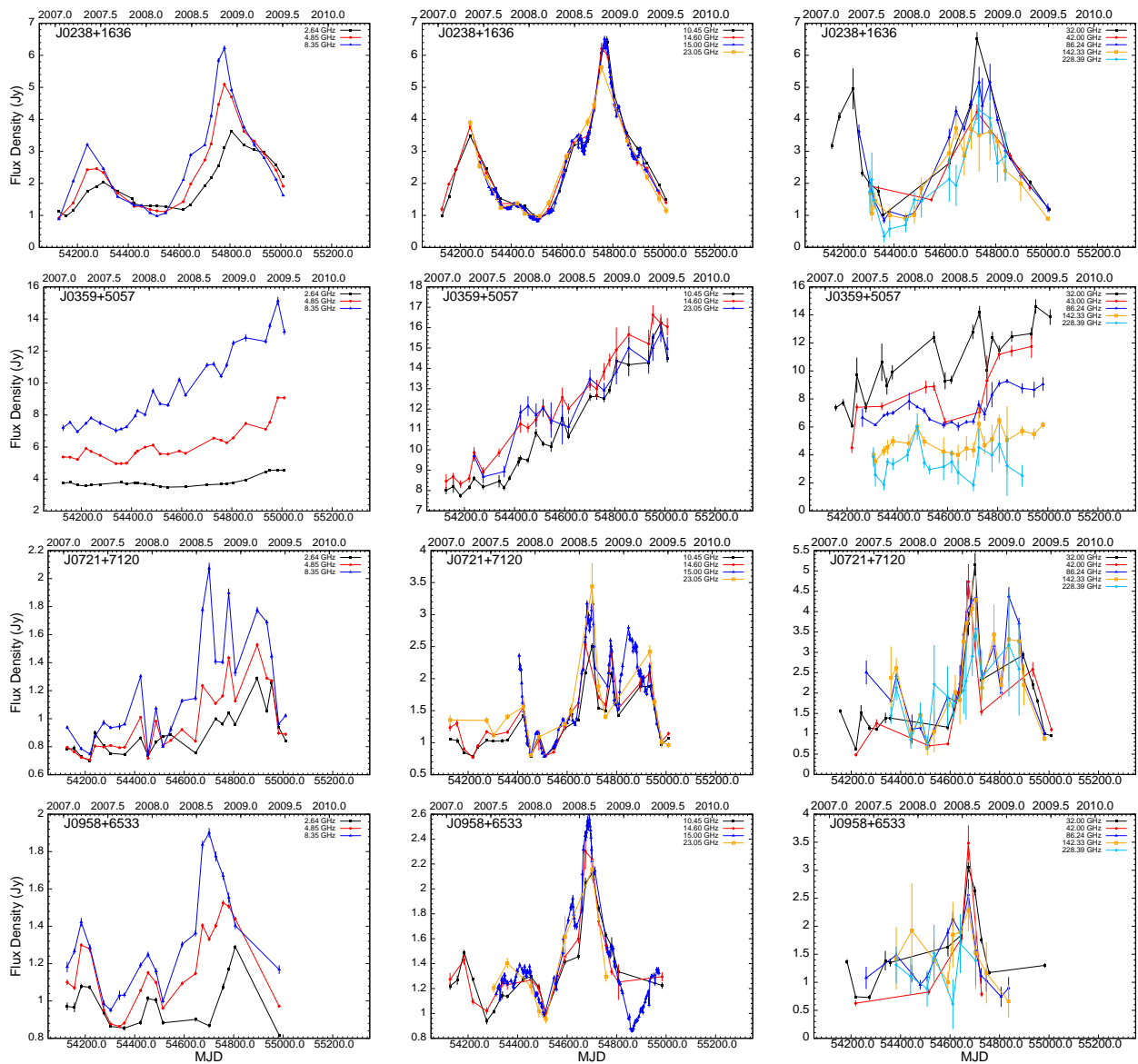}
  \caption{Combined EB and PV light curves for selected, particularly active 
{\it Fermi} 
sources of the F-GAMMA sample. Data obtained within 
the OVRO 40-m program at 15~GHz are also shown. {\bf Left:} low
    frequencies (2.64, 4.85, 8.35~GHz), {\bf Middle:} intermediate frequencies 
(10.45, 14.60/15.00,
    23.05~GHz), {\bf Right:} high frequencies (32.0, 42.0, 86.2, 142.3, 228.9~GHz). From top to bottom are
    shown: J0238+1636 (AO\,0235+16), J0359+5057 (NRAO\,150), J0721+7120 (S5\,0716+714) and J0958+6533
    (S4\,0954+65)}
  \label{LC_comb_all}
\end{figure*}

\begin{figure*}
  \centering
  \includegraphics[angle=0,width=0.9\textwidth]{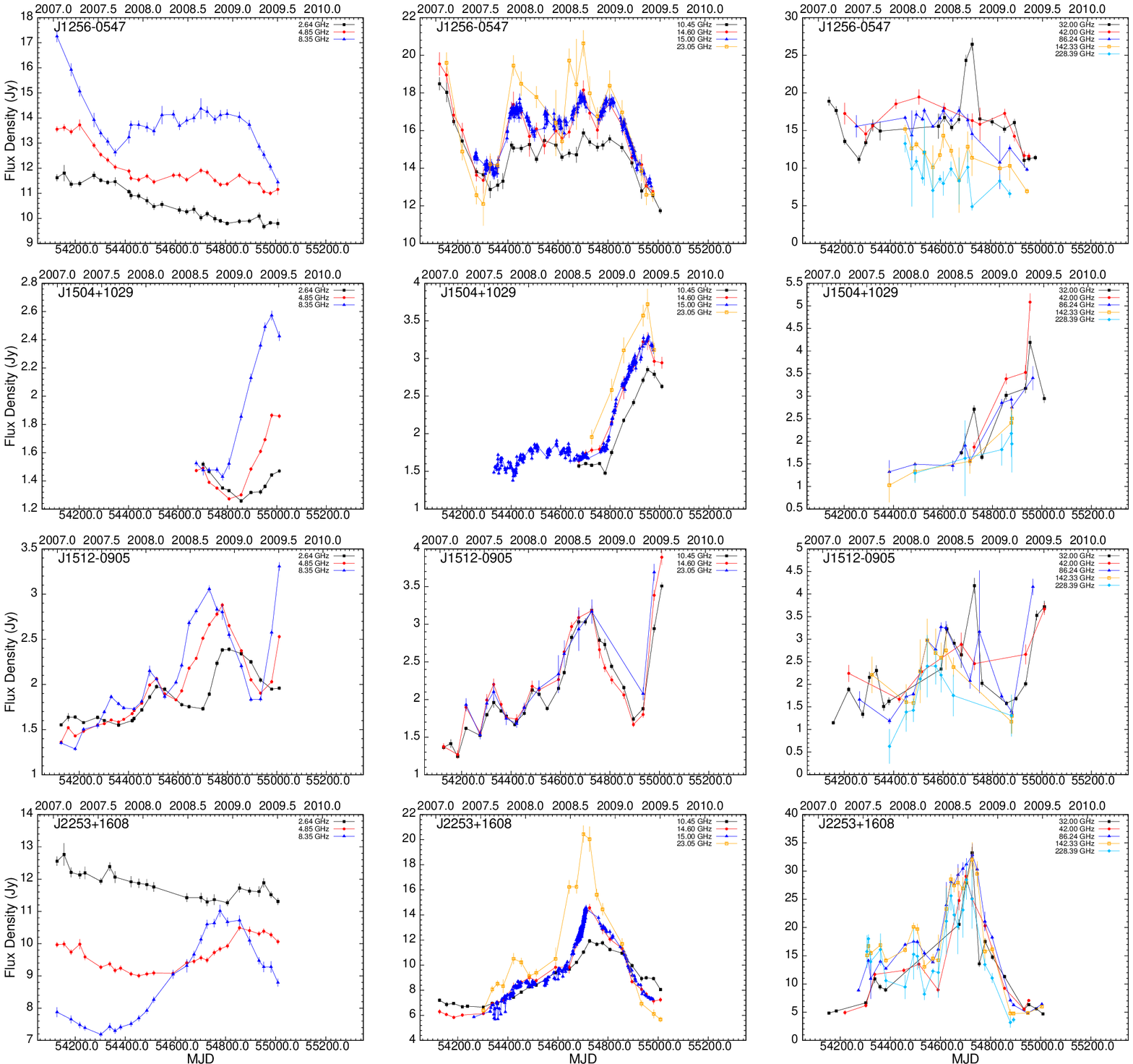}
  \caption{Columns as in Fig.~\ref{LC_comb_all} for four additional sources, 
namely J1256-0547 (3C\,279), J1504+1029 (PKS\,1502+106), J1512-0905 
(PKS\,1510-089) and J2253+1608 (3C\,454.3)}
  \label{LC_comb_all2}
\end{figure*}

Examples of combined EB/PV (2.6 to 228~GHz) light curves of active {\it Fermi} 
$\gamma$-ray sources are shown in
Figs.~\ref{LC_comb_all} and~\ref{LC_comb_all2} demonstrating the variety of cm 
to short-mm band variability
behaviours.
Strong, correlated outbursts across the observing period are visible. Those 
typically last from months to about one year and are seen in nearly all  
bands, often delayed and with lower variability 
amplitudes towards lower frequencies, e.g. in J0238+1636 (AO\,0235+164) or 
J2253+1608 (3C\,454.3). Often more fine structure, i.e. faster variability, is 
seen towards shorter wavelengths. In several cases, like J1256-0547
(3C\,279) and J2253+1608 (3C\,454.3), there is no or very little variability
in the lowest frequency
bands (4.85 and 2.64~GHz). The case of J0359+5057 (NRAO\,150) also demonstrates
the presence of different variability properties with a nearly 
simultaneous, monotonic  total flux density increase at all
frequencies on much slower time scales (years). 
Interestingly,  the latter 
happens without strong spectral changes (see also Sect.~\ref{spec}).

Figures~\ref{LC_comb_all} and~\ref{LC_comb_all2} also include the denser 
sampled light curves obtained by the OVRO program at 15~GHz.  
Even visual comparison 
demonstrates the 
good agreement between the programs. The 
EB data nicely resemble the variability behaviour seen at higher 
time resolution with the OVRO 40-m. To evaluate the agreement of the 
cross-station calibration, at the nearby frequencies of 15 and 14.6 GHz, we 
compared the EB and OVRO data sets. For the 
calibrators (non-variable steep spectrum sources) the difference is about 1.2\% 
and can be accounted for by the source spectrum, given the slightly different 
central frequencies at EB and OVRO. Occasional divergence from this minor effect 
can be explained by pointing offsets and source spectral variability. Generally, 
the two stations agree within $<$3--4\%.

To study the flare characteristics, a detailed light curve analysis was 
performed.
For consistency we used the 2-year data set, from June 2007 to June 2009, where 
quasi-simultaneous data from both, EB and PV are available.
The analysis included (i) a $\chi^{2}$-test for assessing the significance of 
the source variability, (ii) determination of the variability amplitude and 
its frequency dependence and (iii) estimation of the observed flare time 
scales over the considered time period. Subsequently, multi-frequency 
variability brightness temperatures and Doppler factors are calculated. The 
analysis and results are discussed in the following sections. 

%
%

\subsection{Assessing the significance of variability }\label{var_test}
For each source and frequency, the presence of significant variability was examined under the hypothesis of a
constant function and a corresponding $\chi^{2}$ test. A light curve is considered
variable if the $\chi^{2}$ test gives a probability of ${\le} 0.1\%$ for the assumption of constant flux
density (99.9\% significance level for variability).

The $\chi^{2}$ test results reveal that all primary (EB) and secondary (PV) calibrators are non-variable, as
expected. The overall good calibration/gain stability is demonstrated by the small residual (mean) scatter
($\Delta S/\left< S \right>$) in the calibrator light curves of 0.6 to 2.7\% between 2.6 and 86.2~GHz,
respectively. At 142.3 and 228.9~GHz, however, these values increase by a factor of up to 2--3. Almost all
target sources of the F-GAMMA sample (for which the available data sets at the given frequency were
sufficiently large) are variable across all bands. Taking all frequencies into account, we obtain a mean of
91\% of target sources to be significantly variable at a 99.9\% significance level.  We note a trend of lower
percentage of significantly variable sources (between 78 to 87\%) at 43, 142.3 and 228.9~GHz. At these bands
the measurement uncertainties are significantly higher due to lower system performance and increasing
influence of the atmosphere. Consequently, the larger measurement uncertainties at these bands
reduce our sensitivity in detecting significant variability, particularly in the case of low amplitude
variations and weaker sources. For subsequent analysis, only light curves exhibiting significant variability,
according to the $\chi^{2}$ test, are considered.

\subsection{Dependence of variability amplitudes on frequency} \label{amplitudes}
To quantify the strength of the observed flares in our light curves, we calculated the modulation index
$m_{\mathrm{\nu}}=100\times \sigma_{\mathrm{\nu}}/\left<S_{\mathrm{\nu}}\right>$ , where
$\left<S_{\mathrm{\nu}}\right>$ is the mean flux density of the light curve at the given frequency and
$\sigma_{\mathrm{\nu}}$ its standard deviation.  The calculated modulation
indices for each source and frequency show a clear trend. The sample averaged variability amplitude
$\left<m_{\mathrm{\nu}}\right>$ steadily increases towards higher frequencies from 9.5\% at 2.6~GHz to 30.0\%
at 228.9~GHz. A similar behaviour has been found in previous studies
\citep[e.g.][]{1992A&A...254...71V,2004A&A...419..485C} though for smaller source samples, lower frequency
coverage (typically up to 37~GHz) and different definitions of the variability amplitudes.

In order to establish this trend as an ``intrinsic property'', possible biases in the calculated variability
amplitudes have been taken into account as follows. First, due to the dependence of the mean flux density
$\left<S_{\mathrm{\nu}}\right>$ on frequency, we only consider the standard deviation of the flux density of
each light curve, as a measure of the mean variability/flare strength at a given frequency. Second, redshift
effects are removed by converting to rest-frame frequencies. Finally, we remove the possible influence of
measurement uncertainties (finite and different sampling, single-flux-density uncertainties) by computing
intrinsic values for the flux density standard deviation of each light curve, using a likelihood
analysis. That is, assuming a Gaussian distribution of fluxes in each light curve, we compute the joint
likelihood of all observations as a function of both the intrinsic mean flux density $S_0$ and intrinsic
standard deviation $\sigma_0$, accounting for the different measurement uncertainties $\sigma_j$, at each flux
density measurement $S_j$, and the different number of flux density measurements for different light
curves. The joint likelihood for $N$ observations is \citep[][Eq. 20 with derivation
therein]{2011ApJS..194...29R}:
\begin{eqnarray}
\mathcal{L}(S_0,\sigma_0) &=&  \left( \prod_{j=1}^N
\frac{1}{\sqrt{2\pi(\sigma_0^2 + \sigma_j^2)}}\right)\times \nonumber \\
&&  \exp \left[
-\frac{1}{2}\sum_{j=1}^N\frac{(S_j-S_0)^2}{\sigma_j^2 + 
\sigma_0^2}\right].
\end{eqnarray}
Marginalising out the intrinsic mean, $S_0$, we can obtain the maximum 
likelihood values for $\sigma_0$, as well as uncertainties on this estimate. 
We consider a source to be variable at a given frequency, if
$\sigma_0$ at that frequency is more than three sigma away from 0.
\begin{figure}
 \centering
   \includegraphics[trim=0pt 0pt 25pt 0pt  ,clip,width=0.25\textwidth,angle=-90]{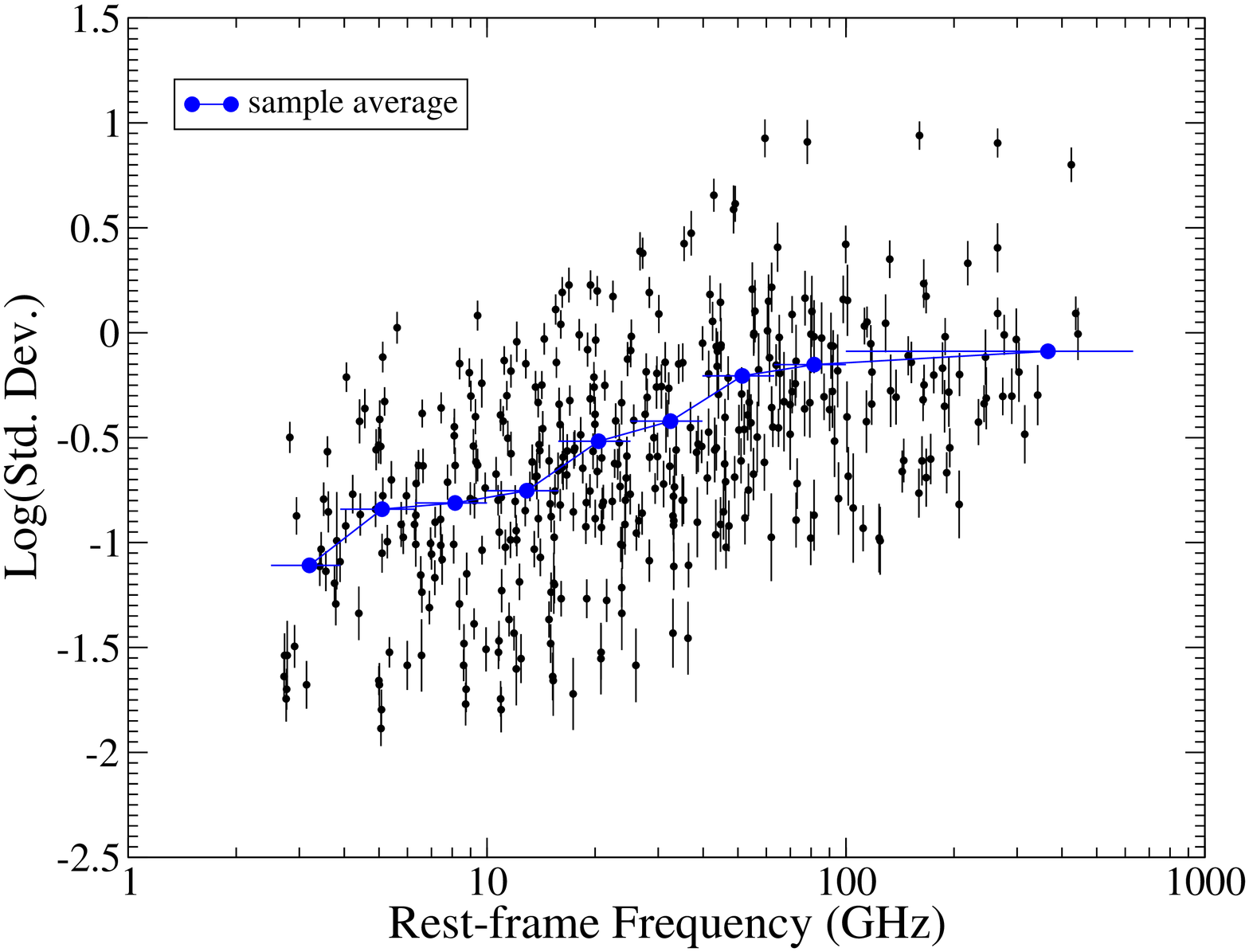}
   \vspace{-.1cm}\\
   \includegraphics[trim=0pt 0pt 27pt 0pt  ,clip,width=0.25\textwidth,angle=-90]{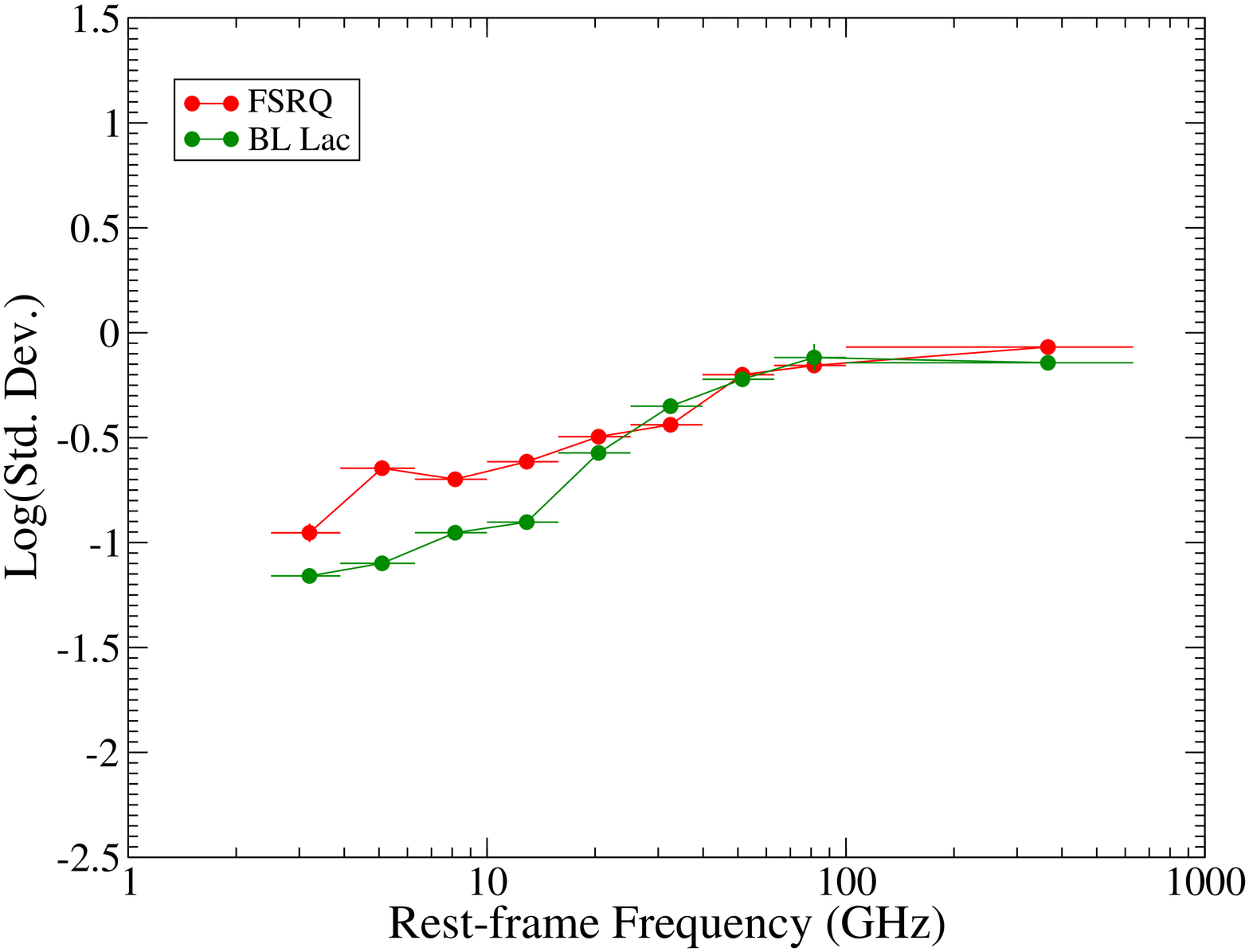}
   \vspace{-.07cm}\\
   \includegraphics[trim=0pt 0pt 0pt 0pt  ,clip,width=0.26\textwidth,angle=-90]{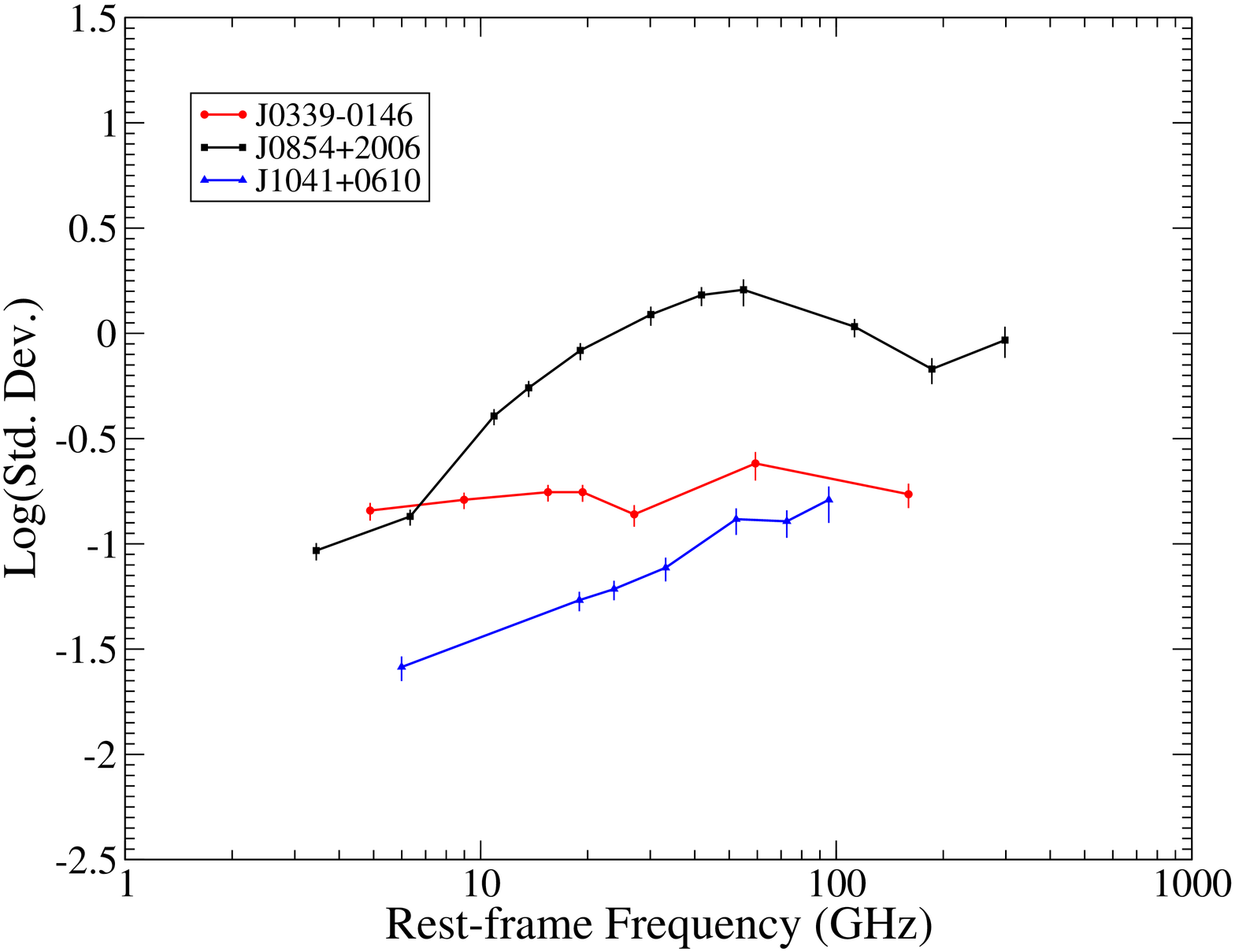}
  \caption{The strength of variability (flux standard deviation) with respect to rest-frame frequency as
    observed with EB and PV. {\bf Top:} scatter plot (black, all sources) showing increasing variations with
    frequency. 
    Superimposed are shown the logarithmic sample averages of the standard deviation after binning in
    frequency space.  {\bf Middle:} logarithmic averages of the standard deviation obtained for the FSRQs
    (red) and BL\,Lacs (green) in the sample.  {\bf Bottom:} three examples of individual sources showing (i)
    a rising trend (blue); (ii) a clear peak (black) and (iii) a nearly flat trend of variability amplitude
    across our bandpass (red).}
  \label{m_vs_nu}
\end{figure}
%


The results are summarised in Fig.~\ref{m_vs_nu} which shows the logarithmic 
flux standard deviations versus
rest-frame frequency for each source, as well as the logarithmic sample 
averages per frequency bin. A clear and significant overall increase 
of the variability amplitude with increasing frequency is observed for the 
F-GAMMA sample, with a maximum reached at rest-frame frequencies 
$\sim$60--80~GHz; i.e. the mm bands. At higher frequencies a plateau or a 
decreasing trend appears to be present but cannot be clearly established due to 
the small number of data points. Looking into individual source patterns, 
however, reveals basically three different behaviours: (i) sources showing only 
a rising trend towards higher frequencies, (ii) sources showing a clear peak at 
rest-frame frequencies between $\sim$40--100~GHz with a subsequent decrease, 
and (iii) few sources showing only a nearly flat trend across our bandpass. 
Typical examples are shown in the bottom panel of Fig.~\ref{m_vs_nu}.  
Consequently, the high frequency plateau in Fig.~\ref{m_vs_nu} (top) is 
the result of averaging over these different behaviours.

Detailed studies of single sources and isolated flares 
\citep[e.g][]{2008A&A...485...51H,2016A&A...590A..48K} 
at their different stages (maxima, slopes of the raising and decaying regimes; 
see Fig.~\ref{m_vs_nu}) are required for detailed comparisons with model 
predictions \citep[e.g.][]{2015A&A...580A..94F}. However, the above findings 
(raising and peaking sources) are qualitatively in good agreement with 
predictions of shock-in-jet models 
\citep[e.g.][]{1985ApJ...298..114M,1992A&A...254...71V,2000A&A...361..850T,
2011A&A...531A..95F,2015A&A...580A..94F}, where the amplitude of flux 
variations is expected to follow three different regimes \citep[growth at high 
frequencies, plateau, decay towards lower frequencies; 
see e.g.][]{1992A&A...254...71V}, according to the three stages of shock 
evolution \citep[Compton, synchrotron, adiabatic loss 
phases;][]{1985ApJ...298..114M}. 
With good frequency coverage up to 375~GHz, 
\citet[][]{1994ApJ...437...91S} interpreted the overall flaring behaviour of 
their 17 sources also in good agreement with the shock scenario with peaking 
variability amplitudes at $\le$90~GHz, similar to our findings. Similar 
results have been reported also by \citet[][]{2008A&A...485...51H} using a large 
number of individual flares, and by \citet{2016A&A...590A..48K} for the 
broadband flare of PKS\,1502+106.

In this framework, Fig.~\ref{m_vs_nu} demonstrates that the F-GAMMA frequency bands largely cover the decay
stage of shocks, but also provide the necessary coverage towards higher frequencies to study the growth and
plateau stages (peaking sources) and thus, test shock-in-jet models in detail
\citep[][]{2012JPhCS.372a2007A,2015A&A...580A..94F}. Here, our APEX monitoring at 345~GHz will add important
information to the initial shock formation/growth phase for a large number of sources. The origin of the
nearly flat trend seen for some sources, however, needs particular investigation. Such behaviour is not easily
understood within the standard shock scenario. The majority of these sources are also found to exhibit a
different spectral behaviour (see Sect.~\ref{spec}).

Figure~\ref{m_vs_nu} (middle) shows 
the behaviour of the 
logarithmic average of the standard deviations with rest-frame frequency, for 
the FSRQs (red) and BL\,Lacs (green) in our sample. 
It is interesting to note that the FSRQs exhibit systematically 
higher variability amplitudes at lower frequencies (in contrast to e.g. the 
OVRO 15~GHz results, see below). However, this difference decreases at 
rest-frame frequencies $\gtrsim$15~GHz and finally disappears at higher 
frequencies ($\gtrsim$20--30~GHz). This behaviour can be understood for 
BL\,Lacs showing stronger self-absorption and exhibiting flares largely
decaying before reaching the lowest frequencies -- in agreement with our 
spectral analysis (see Sect.~\ref{spec}).

We stress that previous studies found opposite behaviour, i.e. BL\,Lacs exhibiting larger variability
amplitudes than FSRQs at comparable frequencies \citep[e.g.][]{2004A&A...419..485C,2011ApJS..194...29R}.  In
contrast to our measure of the variability amplitude (light curve standard deviations), these studies used
the modulation index with the mean flux density as a normalisation factor.  Although useful quantities
for various studies, the normalisation leads only to apparently higher variability amplitudes for BL\,Lacs due
to the frequency-dependent, and on average lower, flux density of BL\,Lacs compared to FSRQs (in our sample by
a factor 3--4).  

\subsection{Flare time scales}\label{time_scales} 

Under the assumption that the emission during outbursts arises from 
causally connected regions, the average rise and decay time of flares (i.e. 
their time scales, $\tau_{\mathrm{var}}$) can constrain the size of the 
emitting region. Combined with the flare amplitude (Sect.~\ref{amplitudes}), 
this information can yield physical parameters such as the brightness 
temperature and Doppler factor.
To obtain $\tau_{\mathrm{var}}$, and given the large number of data sets 
and flares to be analysed, we do not attempt to study individual flares, but 
instead use a structure function analysis \citep{1985ApJ...296...46S}. In 
particular, two different algorithms have been developed aiming at an automated 
estimation of 
flare time scales \citep[see 
also][]{2012A&A...542A.121M}.


The first 
applies a least-squares regression, of the form 
$SF(\tau)=constant\cdot\tau^{\alpha}$, to the
structure function values at time lag $\tau$.
The regression is 
calculated over time lag windows of different size, providing a correlation 
coefficient for each. This method exploits the fact that, for time lags higher 
than the structure function plateau level, the correlation coefficient should 
undergo a monotonic decrease.  Therefore, the observed time scale could be 
defined as the time-lag $\tau_{\mathrm{reg}}$ for which the coefficient 
regression is maximum. However, a change in the $SF(\tau)$ slope may cause a 
decrease in the regression coefficient before the plateau level is reached. 
Given the limited number of data points per light curve, such changes of slope 
often do not reflect a significant variability characteristic of the light 
curve. In order to overcome this problem, the time scale $\tau_{\mathrm{c1}}$ 
was defined as the first structure function maximum at time lags higher than 
$\tau_{\mathrm{reg}}$.

The structure function of a time series whose variability is characterised by a broken power-law spectrum
shows a plateau after which it becomes approximately flat. This fact is exploited by the second algorithm for
the estimation of the time scale. Defining $SF_{\mathrm{plateau}}$ as the structure function value at the
plateau, we calculated $\tau_{\mathrm{c2}}$ as the lower time lag for which 
$SF(\tau)>SF_{\mathrm{plateau}}$.

The use of automated procedures for the estimation of time scales,
considerably speeds it up 
and provides a fully objective and reproducible method. We compared
the time scales obtained automatically with the ones resulting from visual inspection of the structure
function as well as light curve plots for a sample of sources. The agreement between the results is
satisfactory.  If the difference $|\tau_{\mathrm{c1}}-\tau_{\mathrm{c2}}|$ is 
equal to or smaller than the
average sampling of the investigated light curve, we considered the two values as related to the same time
scale, which is then defined as their average. Large discrepancies 
between the results of the two
methods have been considered as strong evidence of multiple time scales in the light curves. In this case, the
two values have been considered separately. Occasionally, the estimated time scale coincides with the maximum
time lag investigated by means of structure function. This occurs in cases where a (long-term) flare is just
observed as a monotonic trend not changing throughout the whole time span of the observations. In these cases,
the estimated time scale must be considered as a lower limit to the true flare time scale.

The values returned by the structure function have been cross-checked by means of a wavelet-based
algorithm for the estimation of time scales, based on the Ricker mother wavelet \citep[a detailed discussion
of this time analysis method can be found in][]{2012A&A...542A.121M}. Given the fundamental differences
between the structure function and wavelet algorithms for the estimation of time scales, the combined use of
these two analysis tools is very effective for testing the reliability of the results. The substantial
agreement between the two methods is demonstrated by a linear regression between their time scale estimates,
which returns a correlation coefficient of ${\sim} 0.8$.

We note that our about monthly cadence hampers the detection and 
investigation of more rapid flares ($\lesssim$\,days to weeks), while the 
limited 2.5 year time span of the current light curves sets an obvious limit to 
the maximum time scales that can be studied.  Furthermore, the estimation of 
meaningful flare time scales at 228.9~GHz is strongly limited by the often much 
lower number of data points (see Figs.~\ref{LC_comb_all} and 
\ref{LC_comb_all2}), large measurement uncertainties, and the reduced number of 
significantly variable sources (see Sect.~\ref{var_test}). Consequently, we 
exclude the 228.9~GHz light curves from the current analysis.

The estimated flares time scales typically range between 80 and 500\,days. The sample mean and median values
obtained at each frequency are given in Table~\ref{tab:time_scales}. A clear trend of faster variability
towards higher frequencies is found with mean values of e.g. 348\,days (median: 350\,days), 294\,days (median:
270\,days) and 273\,days (median: 240\,days) at 2.6, 14.6 and 86.2~GHz, respectively. According to
statistical tests (K-S test), no significant difference between FSRQs and BL\,Lacs is seen.


\begin{table}
  \caption{Mean and median variability parameters for different frequencies and source class.}   
  \label{tab:time_scales}  
  \centering                     
  \begin{tabular}{r l c c c} 
    \hline
    \hline
\mc{1}{c}{$\nu$} & Class\tablefootmark{a} & $\tau_{\mathrm{var}}$\tablefootmark{b} & Log($T_{\mathrm{B}}$) \tablefootmark{c} & 
$\delta_{\mathrm{var,eq}}$\tablefootmark{d}\\
\mc{1}{c}{(GHz)} &  & (d) & (K) & \\
\hline\\
2.64   & ALL        & 348   ~~350   &  13.60 ~~13.11  & 8.8  ~~8.2 \\
       & FSRQs      & 333   ~~345   &  13.77 ~~13.58  & 12.7  11.5 \\
       & BL\,Lacs   & 339   ~~335   &  13.32 ~~12.24  & 4.7  ~~2.9 \\
4.85   & ALL        & 342   ~~336   &  13.33 ~~12.90  & 7.6  ~~7.5 \\
       & FSRQs      & 355   ~~338   &  13.27 ~~13.05  & 9.3  ~~9.0 \\ 
       & BL\,Lacs   & 325   ~~301   &  13.25 ~~12.09  & 4.4  ~~3.4 \\
8.35   & ALL        & 336   ~~345   &  12.91 ~~12.44  & 5.9  ~~5.8 \\
       & FSRQs      & 331   ~~360   &  12.93 ~~12.72  & 7.6  ~~7.7 \\
       & BL\,Lacs   & 343   ~~345   &  12.77 ~~11.57  & 3.1  ~~2.4 \\
10.45  & ALL        & 327   ~~315   &  12.73 ~~12.34  & 5.3  ~~5.2 \\
       & FSRQs      & 307   ~~278   &  12.80 ~~12.64  & 7.0  ~~6.8 \\ 
       & BL\,Lacs   & 368   ~~353   &  12.45 ~~11.41  & 2.7  ~~2.2 \\
14.60  & ALL        & 294   ~~270   &  12.57 ~~12.20  & 4.8  ~~4.5 \\
       & FSRQs      & 284   ~~263   &  12.67 ~~12.44  & 6.4  ~~6.5 \\
       & BL\,Lacs   & 300   ~~270   &  12.38 ~~11.57  & 2.8  ~~2.5 \\
23.02  & ALL        & 301   ~~270   &  12.37 ~~11.93  & 4.3  ~~4.0 \\
       & FSRQs      & 277   ~~260   &  12.51 ~~12.25  & 5.6  ~~5.5 \\ 
       & BL\,Lacs   & 298   ~~293   &  12.11 ~~11.67  & 2.8  ~~2.6 \\
32.00  & ALL        & 282   ~~280   &  12.38 ~~11.86  & 4.4  ~~4.0 \\
       & FSRQs      & 269   ~~270   &  12.52 ~~12.30  & 5.6  ~~5.7 \\ 
       & BL\,Lacs   & 284   ~~284   &  11.79 ~~11.41  & 2.3  ~~2.3 \\
43.00  & ALL        & 309   ~~300   &  12.12 ~~11.64  & 3.6  ~~3.2 \\
       & FSRQs      & 303   ~~308   &  12.24 ~~11.65  & 4.2  ~~3.7 \\
       & BL\,Lacs   & 294   ~~280   &  11.79 ~~11.39  & 2.0  ~~2.6 \\
86.20  & ALL        & 273   ~~240   &  11.45 ~~11.00  & 2.3  ~~2.2 \\
       & FSRQs      & 261   ~~240   &  11.61 ~~11.09  & 2.9  ~~2.5 \\
       & BL\,Lacs   & 276   ~~240   &  11.02 ~~10.75  & 1.4  ~~1.4 \\
142.30 & ALL        & 328   ~~308   &  11.29 ~~10.62  & 1.9  ~~1.5 \\
       & FSRQs      & 296   ~~300   &  11.50 ~~10.73  & 2.4  ~~1.7 \\
       & BL\,Lacs   & 330   ~~308   &  10.66 ~~10.71  & 1.3  ~~1.3 \\\\
\hline
\end{tabular}
\tablefoot{\tablefoottext{a}{Source class: FSRQs=flat-spectrum radio quasar, BL Lac=BL\,Lac object, ALL=all sources
    including unclassified blazars and radio galaxies};
  \tablefoottext{b}{Variability time scale};\tablefoottext{c}{Logarithm of the variability brutishness
    temperature};\tablefoottext{d}{Variability Doppler factor.}
}
\end{table}

%
\begin{figure}
 \centering
  \includegraphics[trim=90pt 30pt 20pt 70pt  ,clip,width=0.26\textwidth,angle=-90]{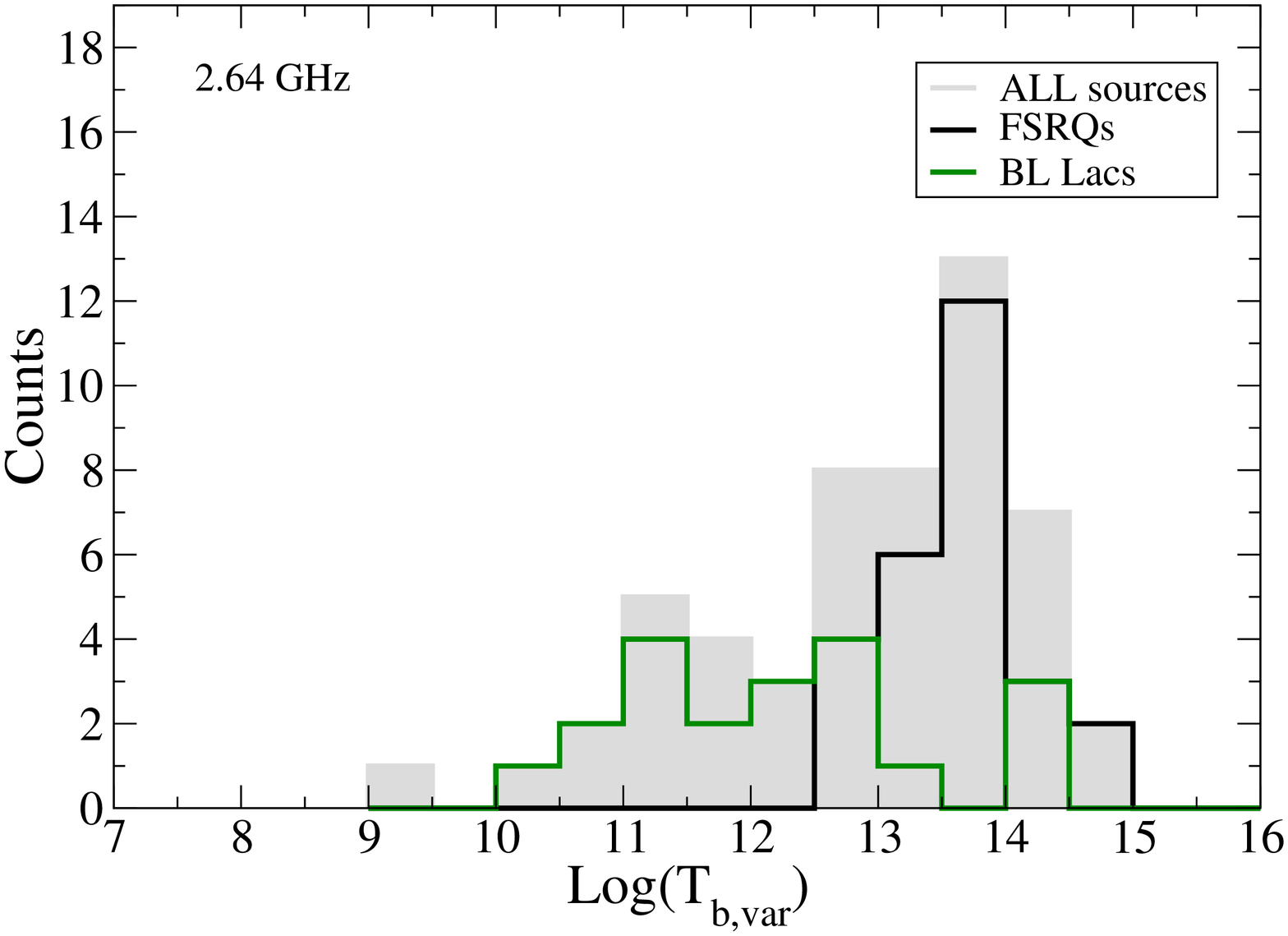}
   \vspace{-.45cm}\\
  \includegraphics[trim=90pt 30pt 20pt 70pt  ,clip,width=0.26\textwidth,angle=-90]{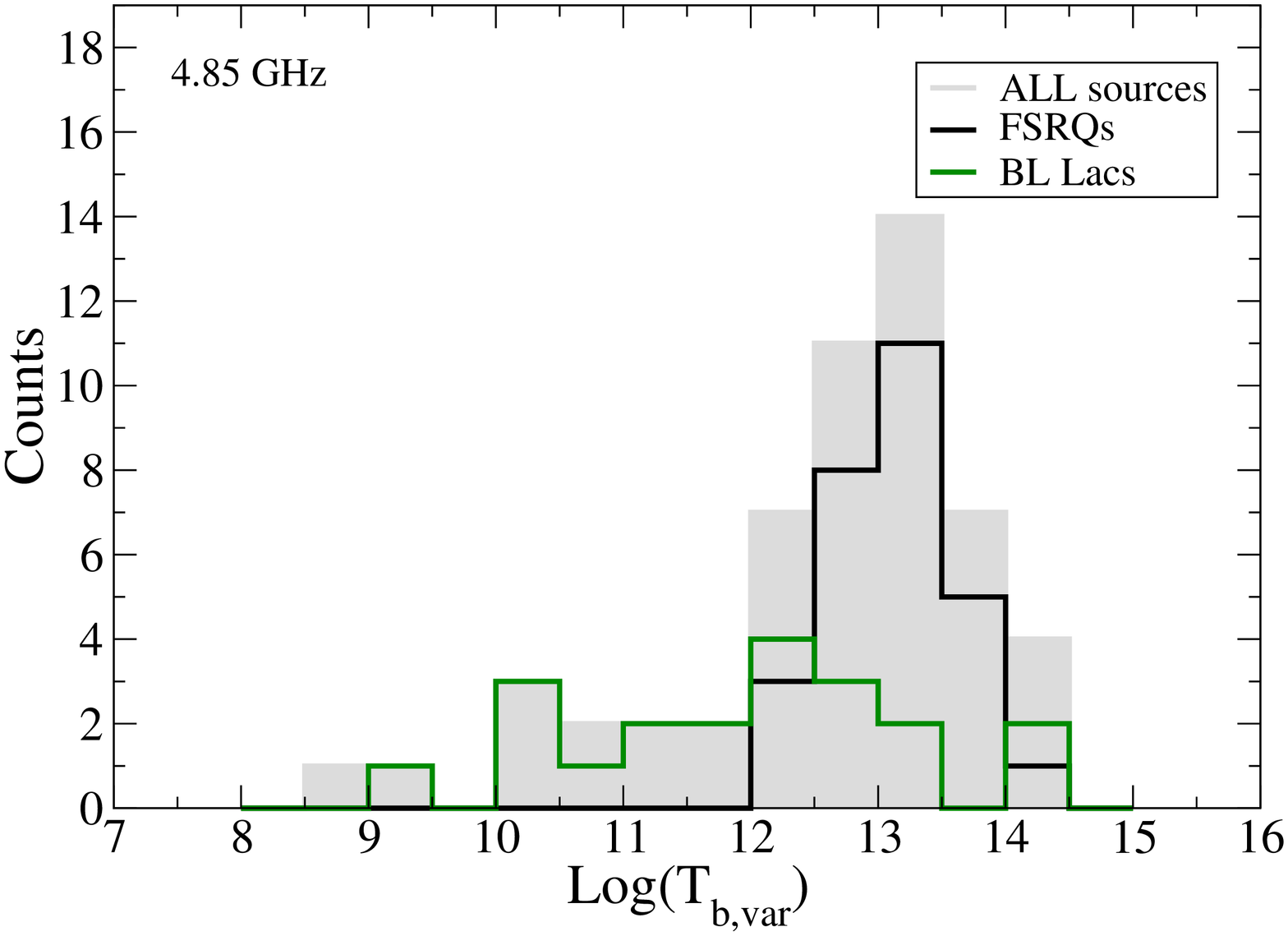}
   \vspace{-.45cm}\\
   \includegraphics[trim=90pt 30pt 20pt 70pt  ,clip,width=0.26\textwidth,angle=-90]{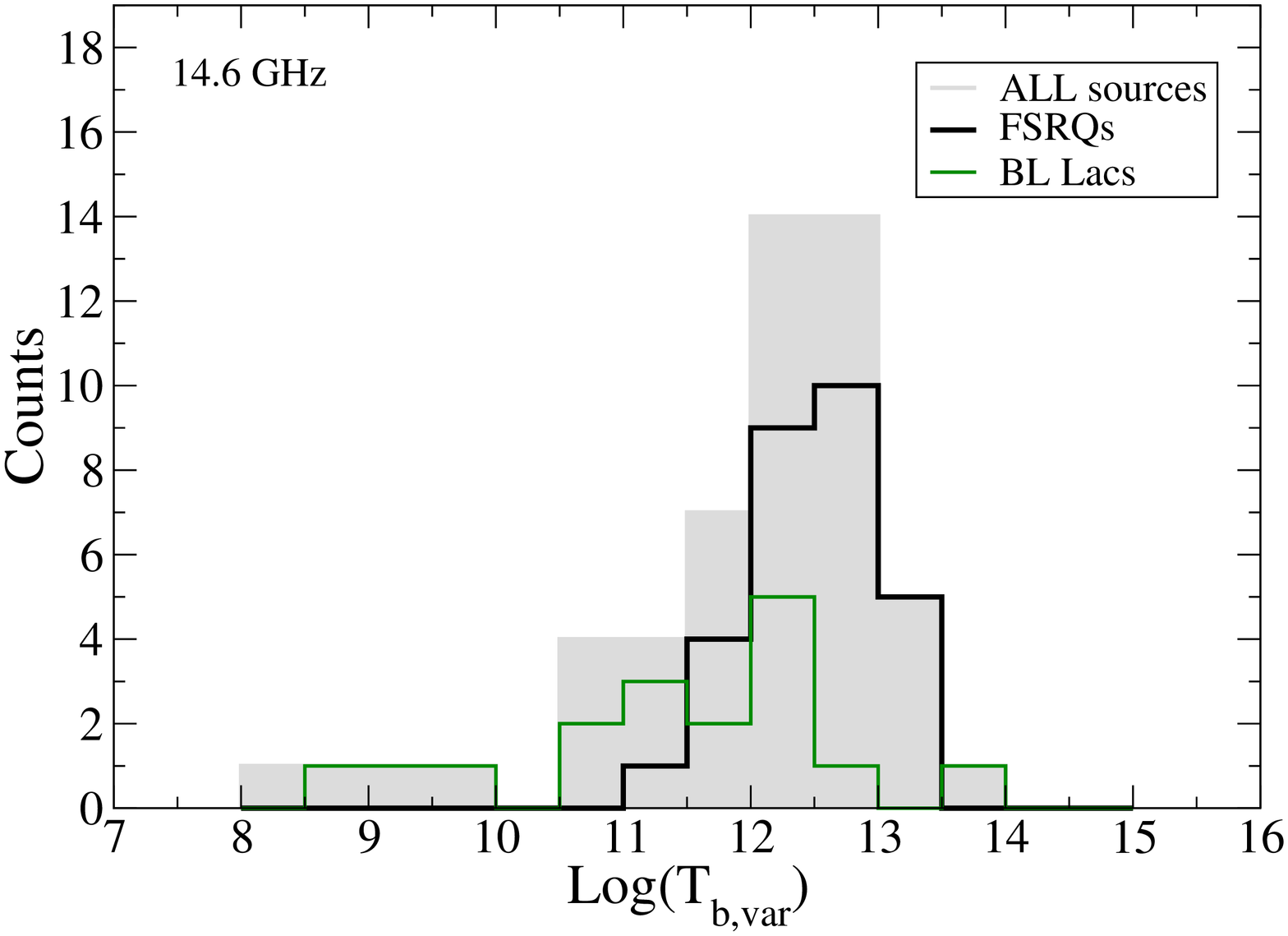}
   \vspace{-.45cm}\\
   \includegraphics[trim=90pt 30pt 20pt 70pt  ,clip,width=0.26\textwidth,angle=-90]{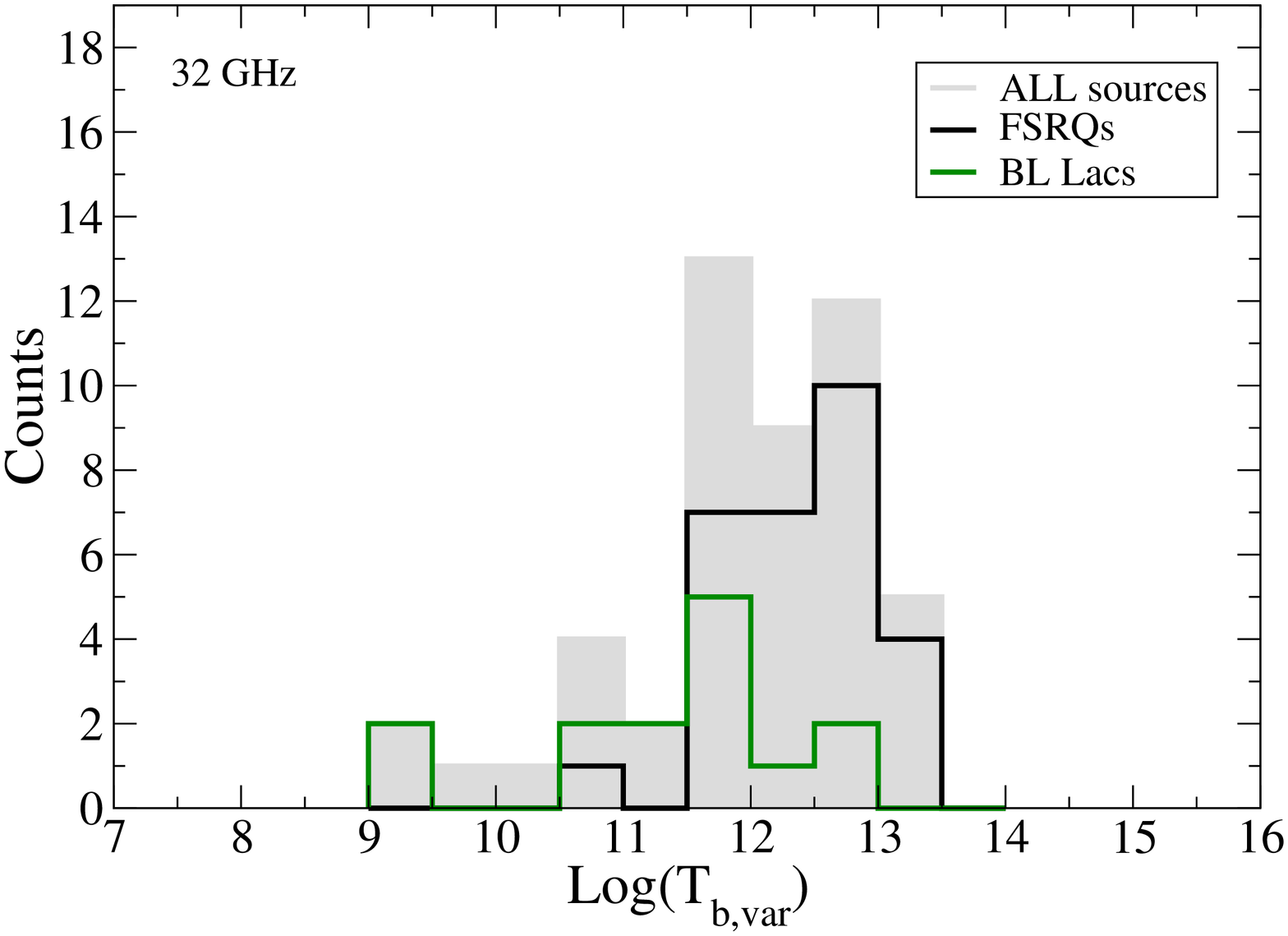}
   \vspace{-.45cm}\\
   \includegraphics[trim=90pt 30pt 20pt 70pt  ,clip,width=0.26\textwidth,angle=-90]{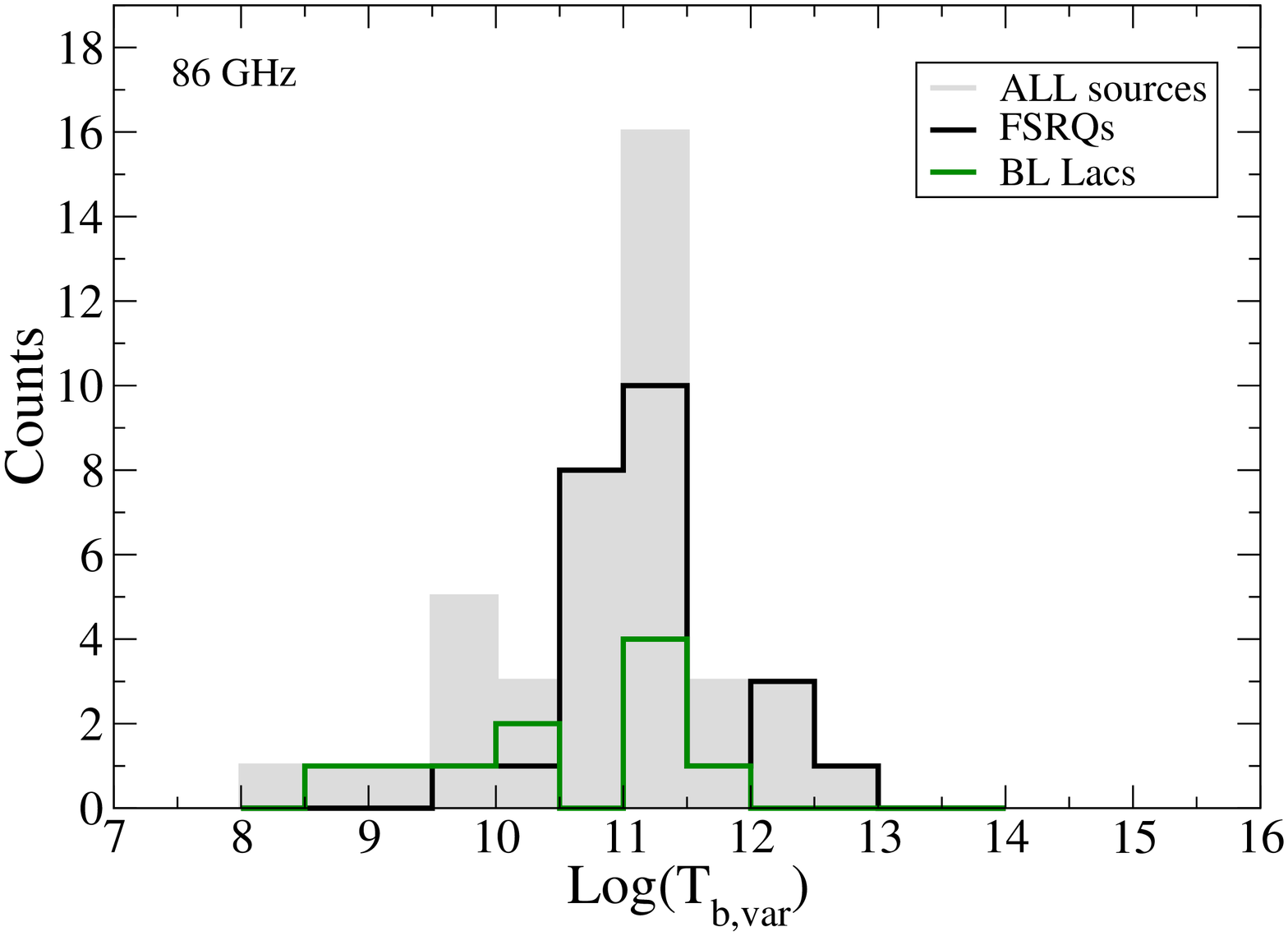}
   \caption{Distributions of apparent variability brightness temperatures at selected frequencies: 2.64, 4.85,
     14.6, 32 and 86~GHz from top to bottom, respectively. All sources (grey) are shown with FSRQs (black) and
     BL\,Lacs (green) superimposed. Note the systematic decrease of $T_{\mathrm{b,var}}$ towards higher
     frequencies as well as the difference in the sample between FSRQs and BL\,Lacs (see text).}
  \label{tb_comb}
\end{figure}

\subsection{Variability brightness temperatures}\label{tb}
Based on the variability parameters discussed above we obtain 
estimates of the emitting source sizes (via the light travel-time argument) and 
the variability brightness temperatures.
Assuming a single emitting component with 
Gaussian brightness distribution,
and given the redshift $z$, luminosity distance $D_{\mathrm{l}}$, the estimated  
time scale $\tau_{\mathrm{var}}$ and amplitude $\Delta S$ ($\sim 
\sqrt{0.5\cdot\rm{SF}\,[\tau]}$\,) of variation, the apparent 
brightness temperature at frequency $\nu$ can be estimated as 
\citep[e.g.][]{2015A&A...575A..55A}

\begin{equation}
T_{\mathrm{b,var}}\mathrm{[K]}=1.47\cdot10^{13}\Delta 
S\mathrm{[Jy]}\,\bigg[\frac{D_{\mathrm{l}}\,\mathrm{[Mpc]}}{\nu\,\mathrm{[GHz]}
~\tau_{\mathrm{var}}\,\mathrm{[days]}~(1+z)^2}\bigg]^2.\label{tb1}
\end{equation}

The calculated values typically range between $10^{9}$ and $10^{14}$\,K. The 
distribution of $T_{\mathrm{b,var}}$ at different frequencies allows, for the 
first time, a detailed study of the frequency dependence of 
$T_{\mathrm{b,var}}$ 
across such a large frequency range. 
Examples of variability brightness temperature distributions for selected 
frequencies are shown in Fig.~\ref{tb_comb}, whereas the sample averaged, 
multi-frequency results are summarised in Table~\ref{tab:time_scales}. We notice 
two main features:
\begin{enumerate}
 \item[(i)] a systematic trend of decreasing $T_{\mathrm{b,var}}$ towards 
higher frequencies, by two orders of magnitude, with mean values of 
$4.0\cdot10^{13}$\,K (median: $1.3\cdot10^{13}$\,K), $3.7\cdot10^{12}$\,K 
(median: $1.6\cdot10^{12}$\,K) and $2.8\cdot10^{11}$\,K (median: 
$1.0\cdot10^{11}$\,K) at 2.6, 14.6 and 86.2~GHz, respectively.

 \item[(ii)] a difference between FSRQs and BL\,Lacs in our sample with a trend 
of higher brightness temperatures for FSRQs as compared to those of BL\,Lacs 
(see Table~\ref{tab:time_scales} and Fig.~\ref{tb_comb}). A significant
difference between the two classes in the sample is statistically confirmed at 
frequencies between 2.6 and 22~GHz by Kolmogorov--Smirnov (KS) tests and 
Student's t-tests, rejecting the null hypothesis of no difference
between the two data sets ($P<0.001$). Towards mm bands, however, this 
difference becomes less significant and vanishes at e.g. 86~GHz (see also 
Fig.~\ref{tb_comb}).
\end{enumerate}

\subsection{Variability Doppler factors}\label{D}
\begin{figure}
 \centering
  \includegraphics[trim=90pt 30pt 20pt 70pt  ,clip,width=0.26\textwidth,angle=-90]{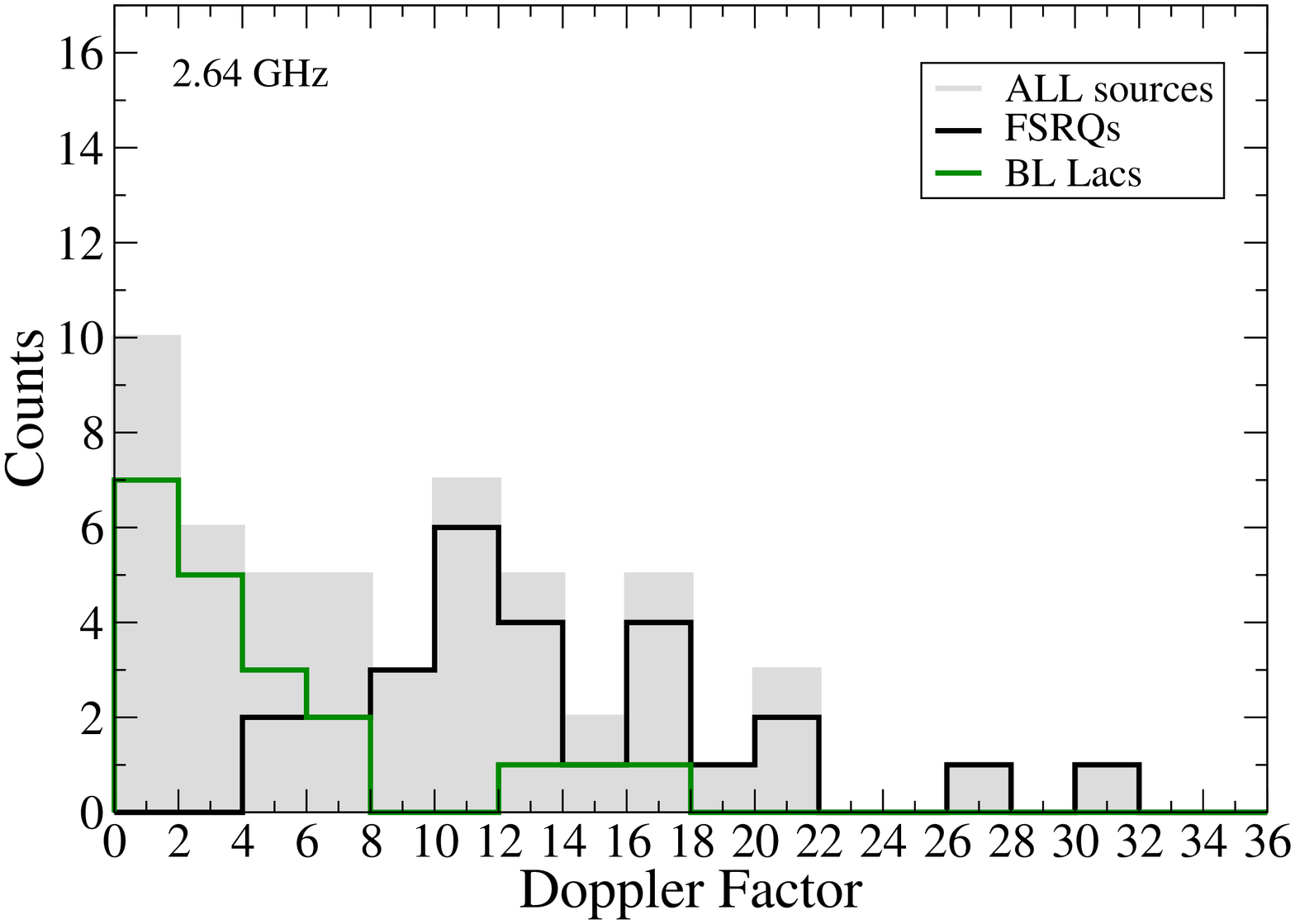}
   \vspace{-.45cm}\\
  \includegraphics[trim=90pt 30pt 20pt 70pt  ,clip,width=0.26\textwidth,angle=-90]{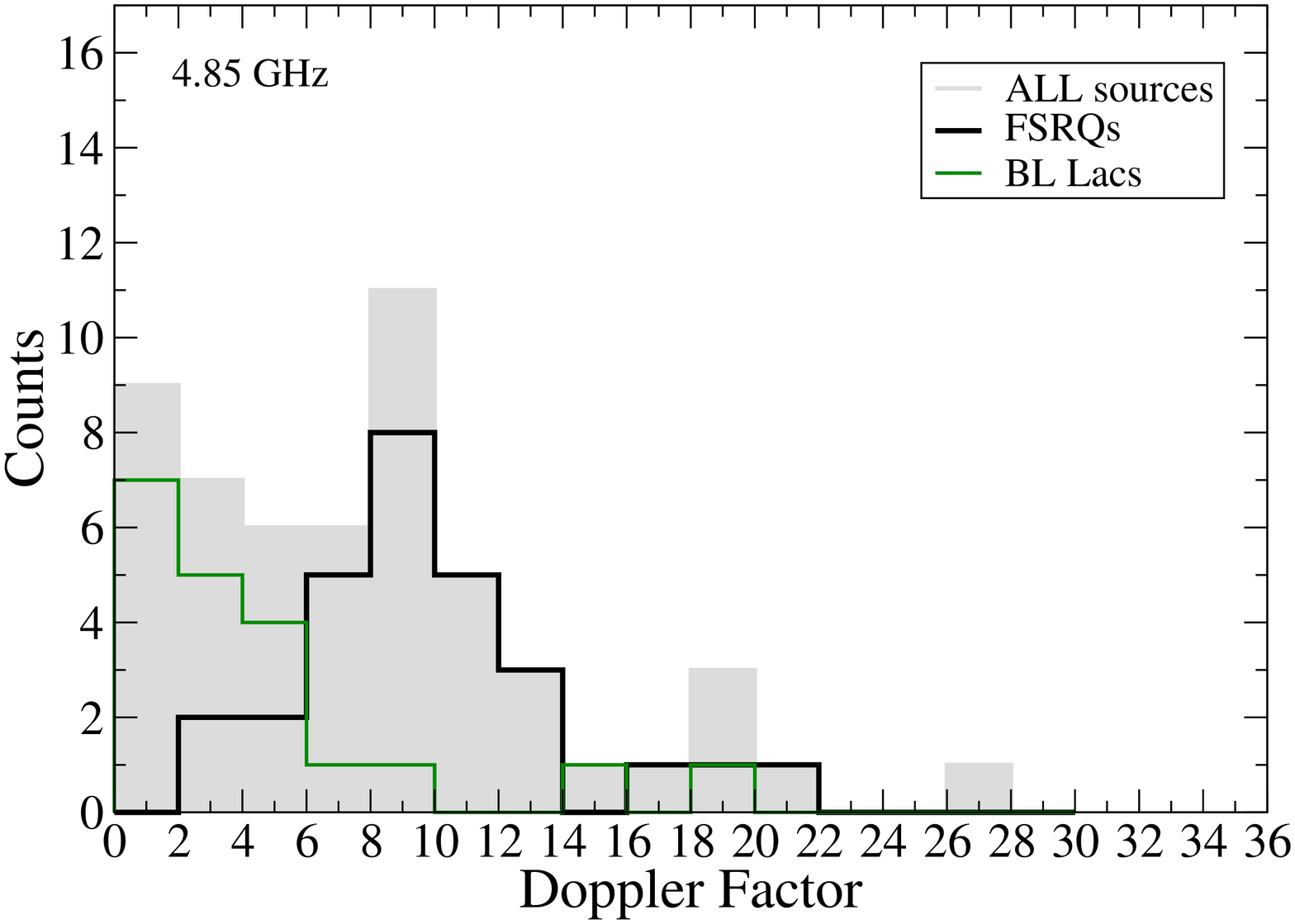}
   \vspace{-.45cm}\\
   \includegraphics[trim=90pt 30pt 20pt 70pt  ,clip,width=0.26\textwidth,angle=-90]{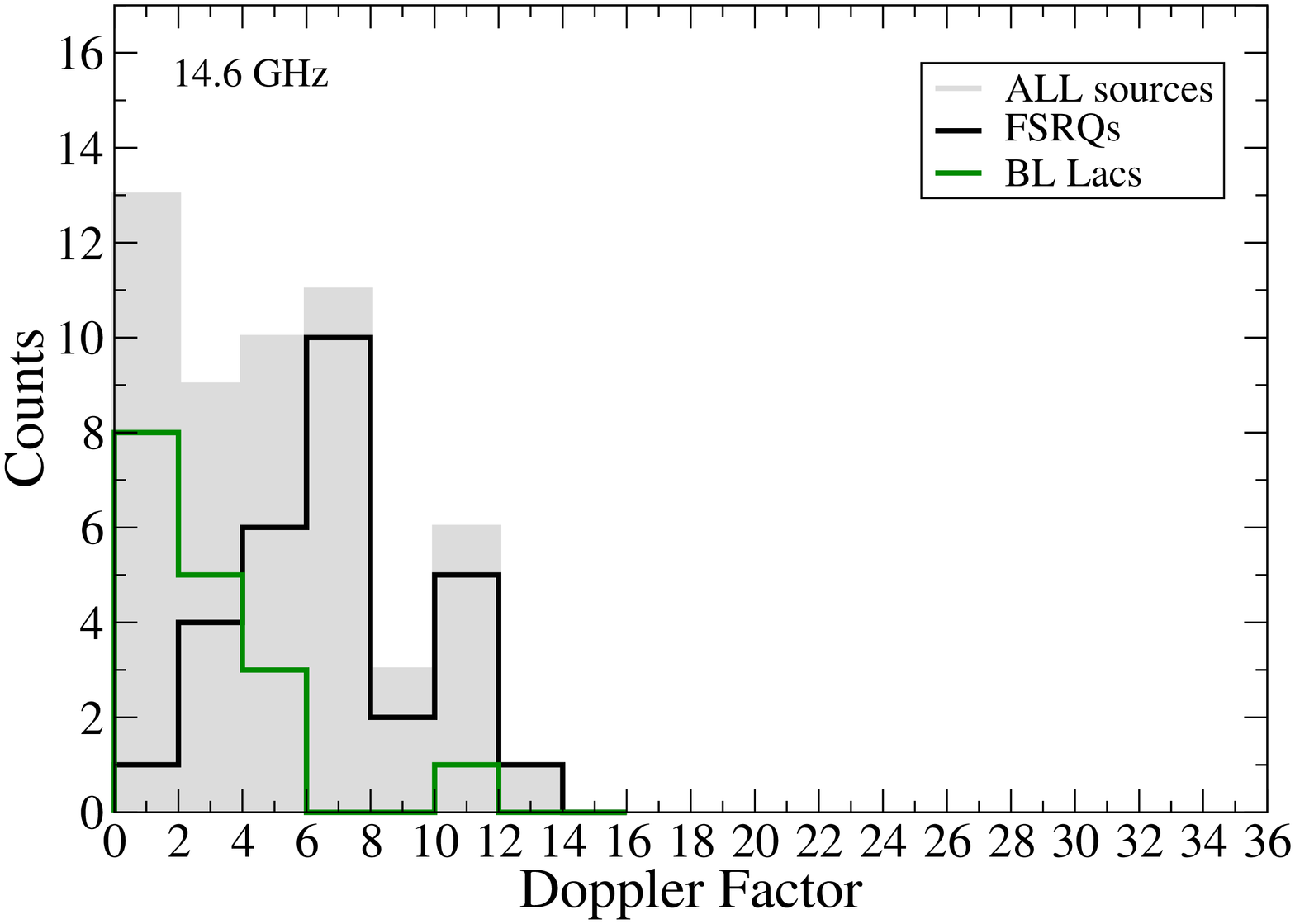}
   \vspace{-.45cm}\\
   \includegraphics[trim=90pt 30pt 20pt 70pt  ,clip,width=0.26\textwidth,angle=-90]{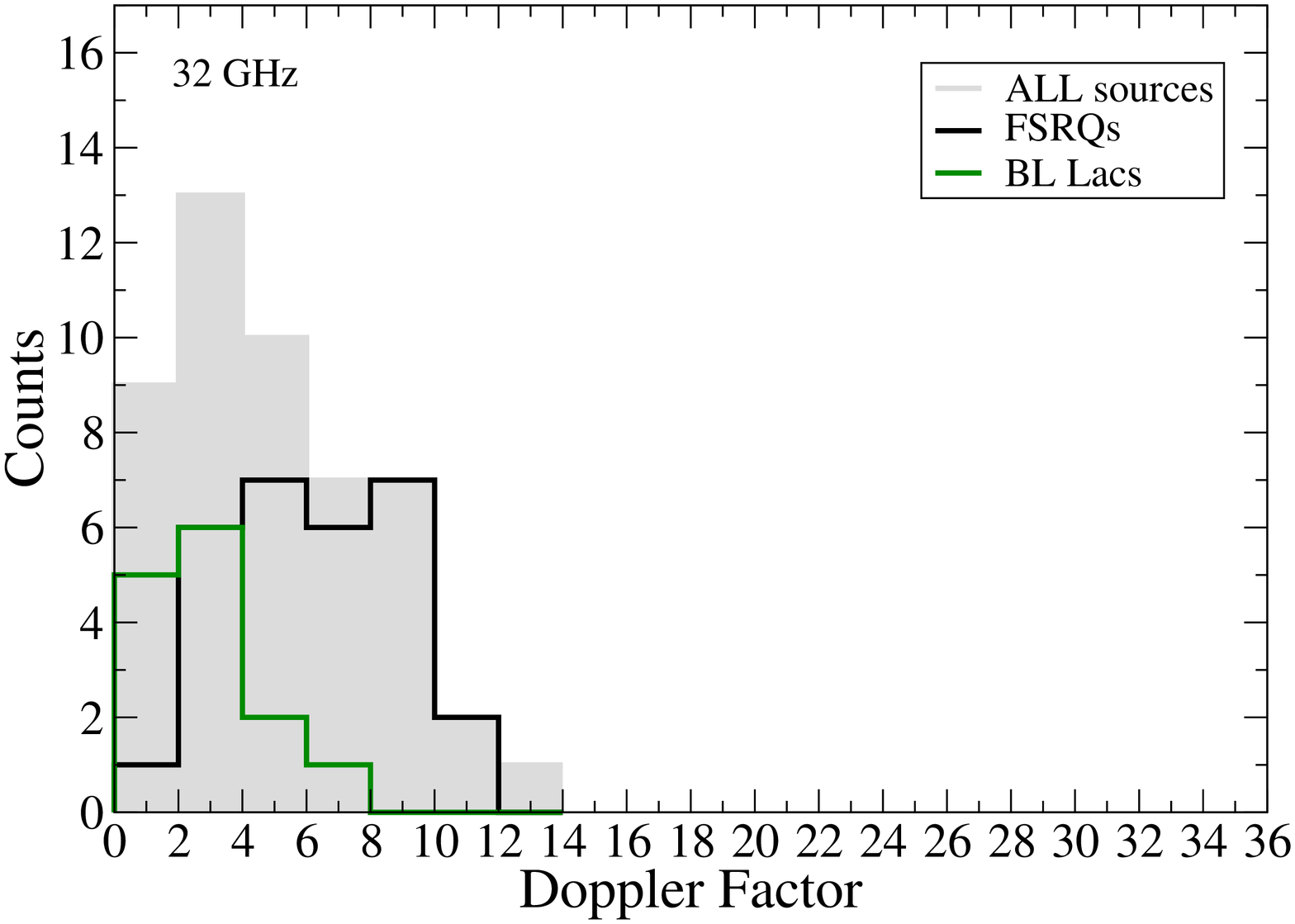}
   \vspace{-.45cm}\\
   \includegraphics[trim=90pt 30pt 20pt 70pt  ,clip,width=0.26\textwidth,angle=-90]{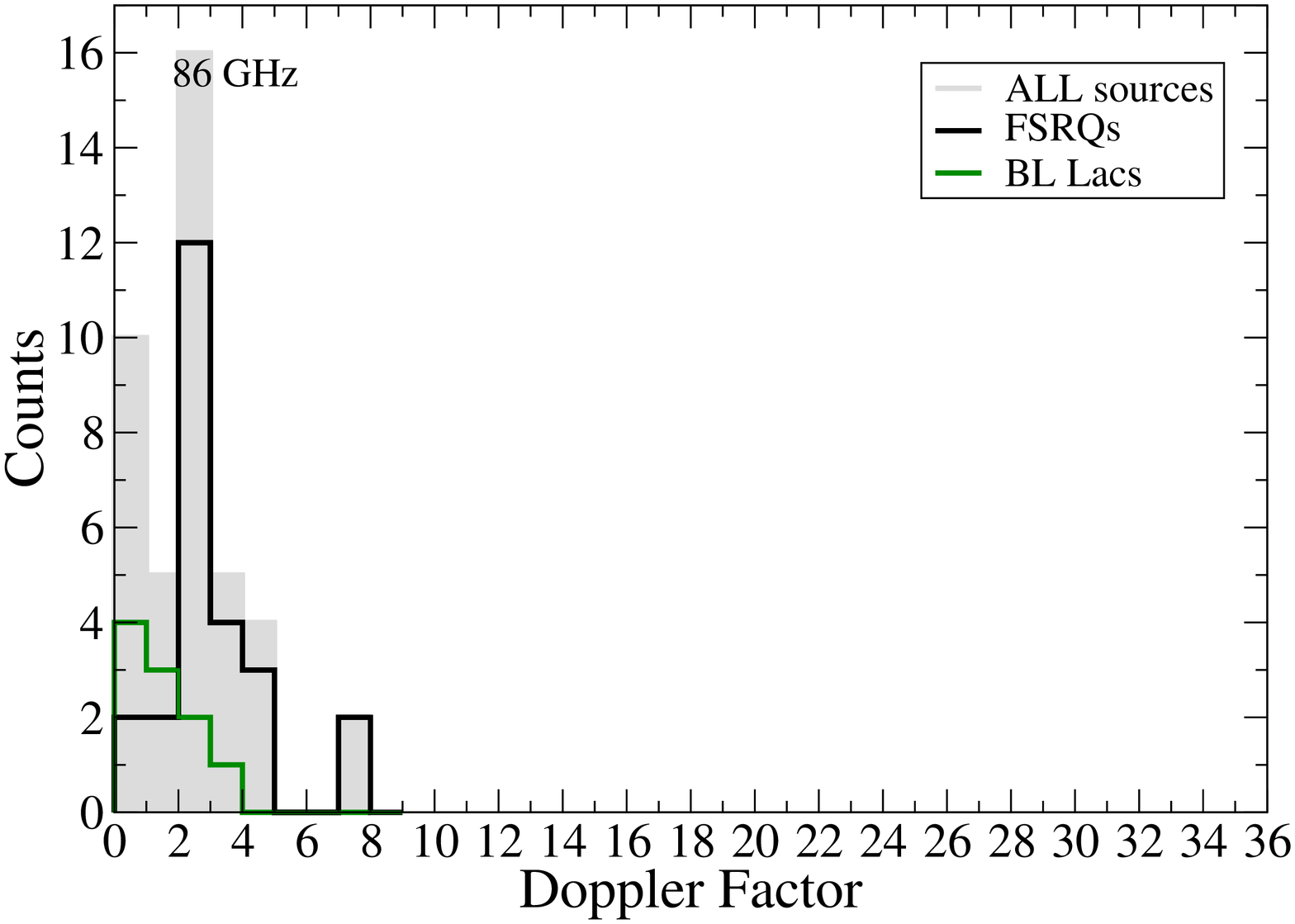}
  \caption{Distributions of variability Doppler factors 
$\delta_{\mathrm{var,eq}}$ at selected
    frequencies: 2.64, 4.85, 14.6, 32 and 86~GHz from top to bottom, respectively. All sources (grey) are
    shown with FSRQs (black) and BL\,Lacs (green) superimposed. Note the systematic decrease of
    $\delta_{\mathrm{var,eq}}$ towards higher frequencies as well as the difference in the sample between
    FSRQs and BL\,Lacs (see text).}
  \label{D_comb}
\end{figure}

%
Under equipartition between particle energy density and magnetic field 
energy density, a limiting brightness temperature, 
$T_{\mathrm{b,int}}^{\mathrm{eq}}$, of $5\cdot10^{10}$\,K is assumed 
\citep[][]{1977MNRAS.180..539S,1994ApJ...426...51R,1999ApJ...511..112L}. We 
estimate the Doppler boosting factors by attributing the excess 
brightness temperature to relativistic boosting of radiation. The 
Doppler factor is then given by 
$\delta_{\mathrm{var,eq}}=(1+z)\sqrt[3-\alpha]{T_{\mathrm{b}}/5\cdot10^{10}}$ 
(with $\alpha\,=\,-0.7$; $S\,\sim\,\nu^{\alpha}$).
The sample averaged Doppler factors at each frequency are summarised in Table 
\ref{tab:time_scales}. Doppler factor distributions at
different frequencies are shown in Fig.~\ref{D_comb}.
%
Two main points can be made:
\begin{enumerate}
 \item[(i)] a systematic trend of decreasing $\delta_{\mathrm{var,eq}}$ towards 
higher frequencies by more than a factor of 4 with mean values of e.g. 8.8 
(median: 8.2), 4.8 (median: 4.5) and 2.3 (median: 2.2) at 2.6, 14.6
and 86.2~GHz, respectively.

 \item[(ii)] a difference between FSRQs and BL\,Lacs in our sample with a trend 
of higher Doppler factors for FSRQs, as compared to those of BL\,Lacs, by a 
factor of 2--3 (see~\ref{tab:time_scales} and Fig.~\ref{D_comb}). Again, a 
significant difference between the two classes in the sample is statistically 
confirmed at frequencies between 2.6 and 22~GHz by KS and Student's t-tests 
rejecting the null hypothesis of no difference between the two datasets 
($P<0.001$). As seen  from Fig.~\ref{D_comb}, the separation of the two 
sub-samples again vanishes towards the highest frequencies confirmed by 
accordingly higher KS test values.
\end{enumerate}

A trend of decreasing $T_\mathrm{b,var}$ and $\delta_{\mathrm{var,eq}}$ 
towards higher frequencies has already been reported  
\citep[e.g.][]{2008A&A...490.1019F, 2016A&A...586A..60K, 2016arXiv160402207L}, 
As pointed out by \citet[][]{1999ApJ...521..493L}, maximum 
(intrinsic) brightness temperatures are expected to occur during the maximum 
development phase of shocks, i.e. at rest-frame frequencies of 
$\sim$\,60--80~GHz (Sect.~\ref{amplitudes}).

Our findings indicate that $T_{\mathrm{b,var}}$ is a decreasing function of
frequency, as expected also from the following considerations.
Starting from Eq.~\ref{tb1} one can write
\begin{equation}
  \label{eq:tb_nu}
  T_{\mathrm{b,var}}\propto \delta S\cdot  \nu^{-2}\cdot  \delta t ^{-2}.
\end{equation}
Assuming that the variability is caused by shocks traveling downstream, $\delta 
S$ is expected to follow an increasing trend with $\nu$ within our bandpass 
\citep{1992A&A...254...71V}. From our dataset (Fig.~\ref{m_vs_nu}) we find that
\begin{equation}
  \delta S \propto \nu^{+0.6}.
\end{equation}
On the other hand ,$\delta t$ is the time needed by the variability event to 
build the amplitude $\delta S$.
This is related to the pace of evolution of the event. Using the mean time 
scales at each frequency for all sources, we find that
\begin{equation}
  \delta t \propto \nu^{-0.1} \Rightarrow \delta t ^{-2}  \propto \nu^{+0.2}.
\end{equation}
If we then substitute $\delta t$ and $\delta S$ in Eq.~\ref{eq:tb_nu} we find that
\begin{equation}
  T_\mathrm{b}\propto \nu^{+0.6} \cdot  \nu^{-2}\cdot  \nu^{+0.2} = \nu^{-1.2}.
\end{equation}
Any divergence from this relation would require an alternative interpretation. 
The logarithm of the median brightness temperatures -- for FSRQs -- shown in 
table~\ref{tab:time_scales}, indeed follows an exponential drop with index 
$-1.17\pm 0.15$ remarkably close to the expected value.
The observed trend may suggest that the Doppler factors of blazars at cm and mm wavelengths are generally
different -- a scenario not unlikely for stratified and optically thick, self-absorbed, bent blazar
jets. While probing a different jet region at each wavelength, each region may exhibit different Doppler
factors depending on jet speed and/or viewing angle. Looking deeper into the jet towards higher frequencies,
decreasing Doppler boosting would then indicate either
increasing Lorentz factors along the jet or jet bending towards the observer for outward motion (with physical
jet acceleration being statistically more likely).
Such interpretation would be supported by the increasing evidence that individual VLBI components of blazar
jets often show changes in apparent jet speed \citep[e.g.][]{2005AJ....130.1418J} with a significant fraction
of these changes being due to changes in intrinsic speed \citep[][]{2009ApJ...706.1253H}.
In the following, we briefly explore and discuss a few alternative explanations.
For instance, another
possibility to explain the observed $T_{\mathrm{b,var}}$ trend is an equipartition brightness temperature
limit changing along the jet: $T_{\mathrm{b,int}}^{\mathrm{eq}}$ might be different at different
frequencies. In order to maintain constant Doppler boosting along the jet, for instance,
$T_{\mathrm{b,int}}^{\mathrm{eq}}$ should decrease towards higher frequencies.  
It must be noted that the relatively lower time sampling at higher frequencies could underestimate the
variability time scales.

The observed trend of higher $T_{\mathrm{b,var}}$ and stronger Doppler boosting 
in FSRQs, as compared to BL\,Lacs, confirms earlier results  
\citep[e.g.][]{1999ApJ...521..493L,2008A&A...485...51H}, however, here 
we demonstrate this effect at a much broader frequency range. This trend is 
furthermore in line with previous VLBI findings of BL\,Lacs exhibiting much 
slower apparent jet speeds and Lorentz factors as compared to those of FSRQs 
\citep[e.g.][]{2010ApJ...723.1150P}. The decreasing difference between FSRQs and 
BL\,Lacs ($T_{\mathrm{b,var}}$ and $\delta_{\mathrm{var,eq}}$) towards higher 
frequencies, however, is in good agreement with the findings of 
Sect.~\ref{amplitudes} and~\ref{time_scales}; the higher variability 
amplitudes of FSRQs at lower frequencies result in correspondingly higher 
brightness temperatures and Doppler factors towards lower frequencies.

\subsection{Spectral variability}\label{spec}

The data obtained at EB and PV are combined to produce about monthly, broadband spectra for each source. The
maximum separation of measurements in a single spectrum is kept below 10 days. This span was chosen as a
compromise between good simultaneity and maximum number of combined EB/PV spectra. Specific sources may show
detectable evolution already beyond 10 days.

\begin{figure*}
  \centering
  \includegraphics[angle=0,width=0.7\textwidth]{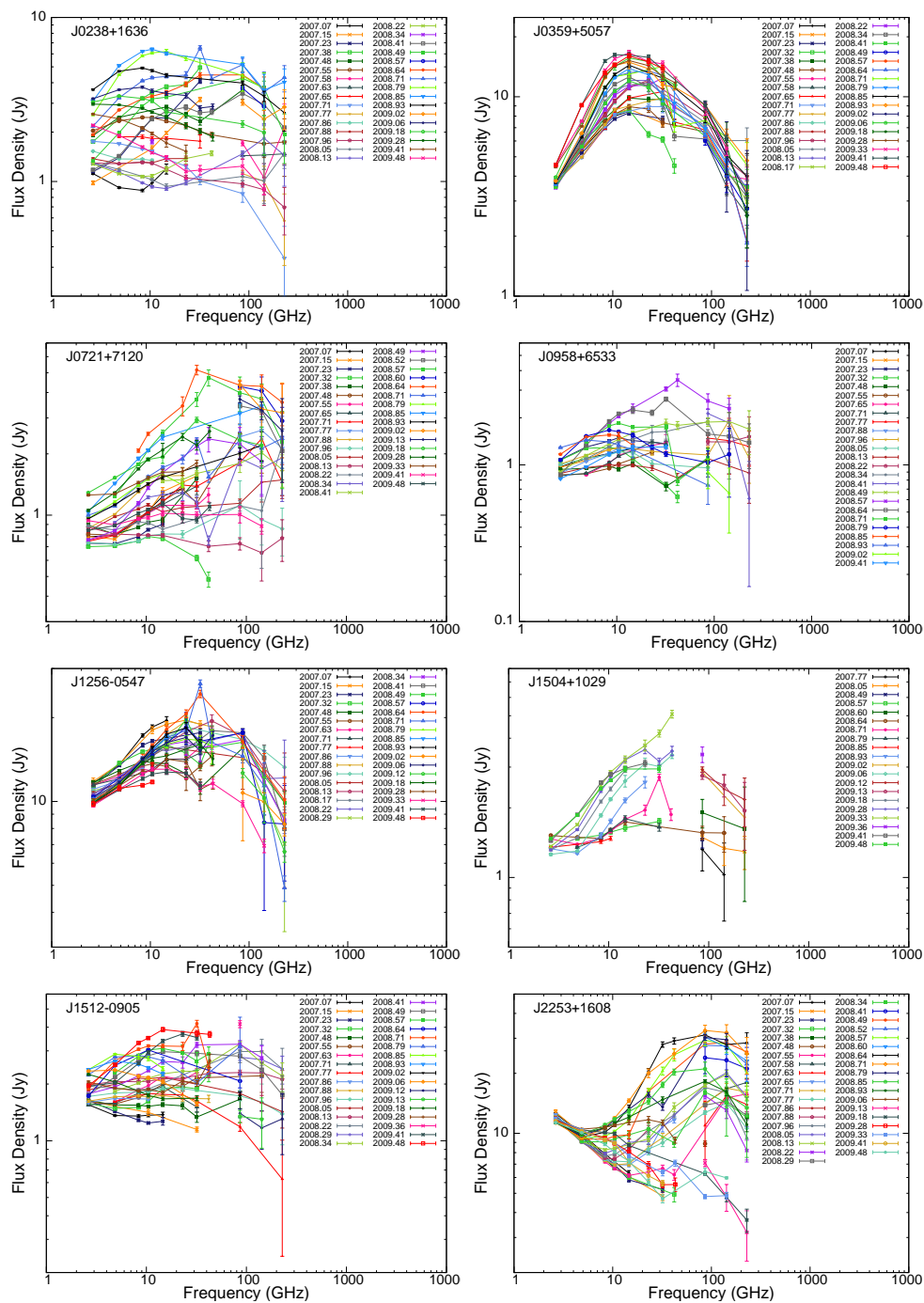}
  \caption{Broadband radio spectra and spectral evolution of the sources shown in
    Figs.~\ref{LC_comb_all} and~\ref{LC_comb_all2} combining the quasi-simultaneous multi-frequency data
    obtained at EB and PV. \cite{2011arXiv1111.6992A, 2012JPhCS.372a2007A} suggests a unification scheme for
    the variability patterns and a simple model that can reproduce all observed phenomenologies.}
  \label{Spec_comb_all}
\end{figure*}

In general, the sources 
show a variety of behaviours. The spectra of the example sources shown in Fig.~\ref{LC_comb_all} and
\ref{LC_comb_all2} are presented in Fig.~\ref{Spec_comb_all}, giving a flavour of the different spectral
behaviours observed. Often the flares seen in the light curves are accompanied by clear spectral
evolution. Their spectral peaks $\nu_{\mathrm{peak}}$ occur first at the highest frequencies and successively
evolve towards lower frequencies and lower flux densities. Typically an evolving synchrotron self-absorbed
component populating a low-frequency steep-spectrum component and unchanged component (quiescent, large-scale
jet) is seen. However, the dominance and broadness of the steep-spectrum component, the lowest frequency
reached by the flare component, and the relative strength of the two components differ from source to source.

The evolution of the flare components appears consistent with the predictions of the shock-in-jet model. That
is, following the three-stage evolutionary path with Compton, Synchrotron and Adiabatic loss phases in the
``$S_{\nu_{\mathrm{peak}}}$--$\nu_{\mathrm{peak}}$'' plane. 
\cite{2012JPhCS.372a2007A} showed that the plurality of
the observed phenomenologies can be classified into only 5 variability classes. Except for the latter type all
other observed behaviours can be naturally explained with a simple two-component system composed of: (a) a
steep quiescent spectral component from a large scale jet and (b) a time evolving flare component following
the “Shock-in-Jet” evolutionary path as the one described before.

Several cases, imply a different variability mechanism with only minor -- if at all -- spectral evolution that
do not seem to be described by the standard three-stage scenario.
J0359+5057 (NRAO\,150) in Fig.~\ref{Spec_comb_all} is a typical example, where almost no broadband spectral
changes are observed during flux changes. In these cases, modifications of the shock-in-jet model or
alternative variability models are required. Geometrical effects in helical, bent or swinging jets
\citep[e.g.][]{1999A&A...347...30V} can possibly be involved in the observed variability \citep[see also
e.g.][]{2013A&A...553A.107C}. Particularly in the case of NRAO\,150, high-frequency VLBI observations indeed
show the presence of a wobbling jet \citep[][]{2007A&A...476L..17A}.

\subsection{Spectral index distribution}\label{spec_indices}
\begin{figure}
 \centering
   \includegraphics[trim=80pt 30pt 23pt 70pt  ,clip,width=0.3\textwidth,angle=-90]{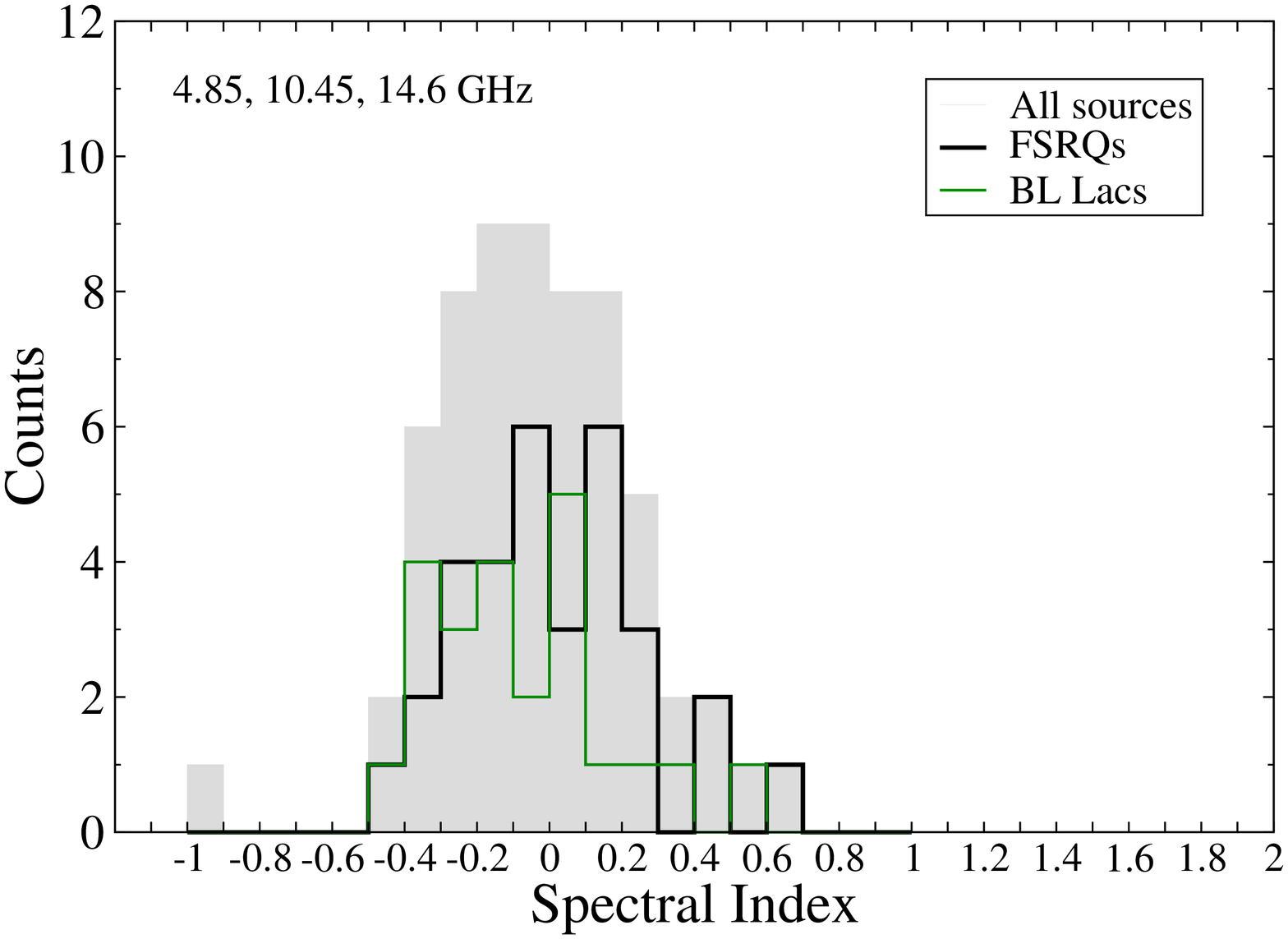}
   \vspace{-.5cm}\\
   \includegraphics[trim=80pt 30pt 20pt 70pt  ,clip,width=0.3\textwidth,angle=-90]{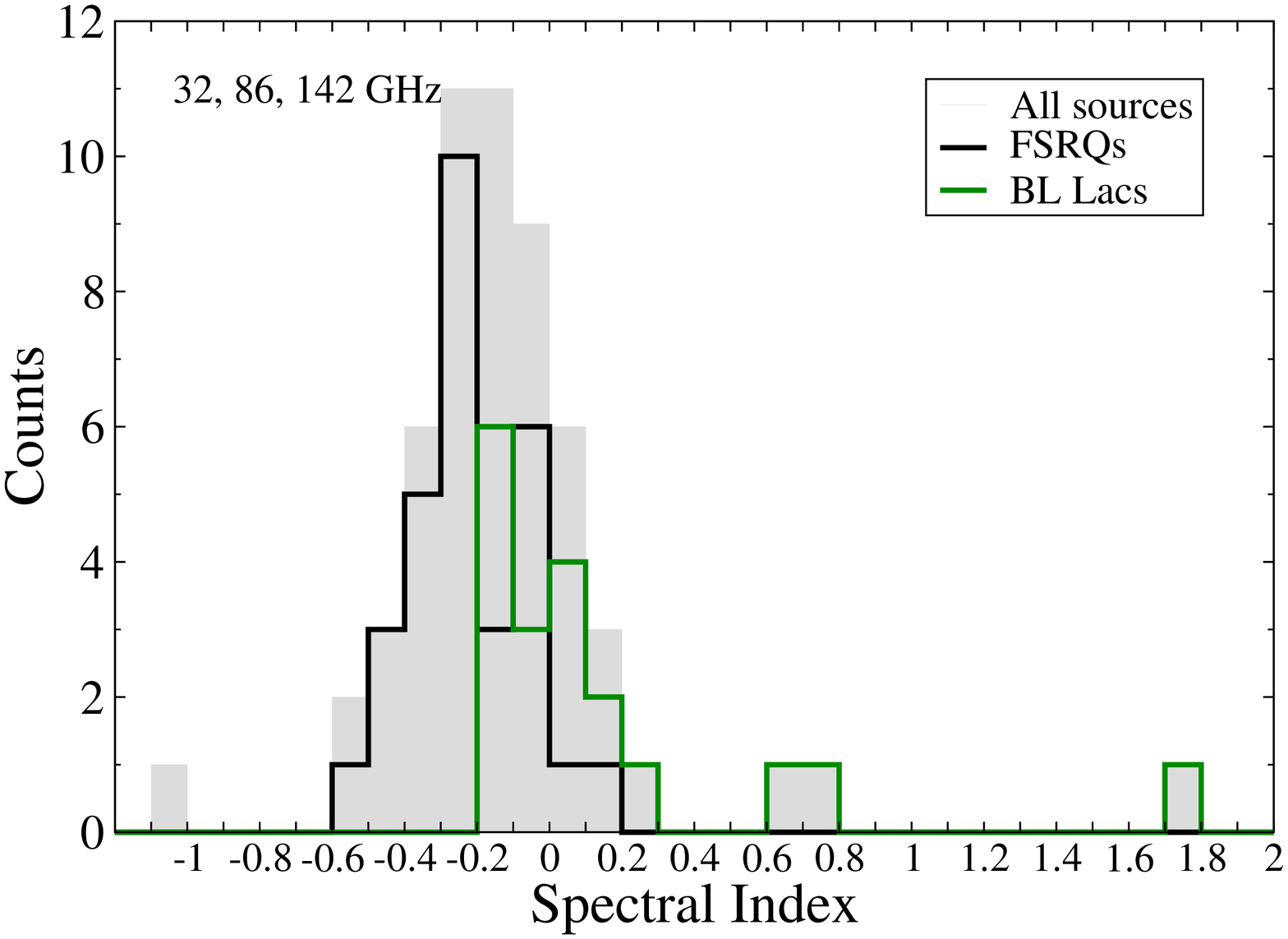}
  \caption{Distributions of the mean spectral index over 2.5 years of
    monitoring: {\bf top:} low-frequency (4.85, 10.45 and 14.6~GHz)
    spectral index, {\bf bottom}: high-frequency (32, 86.2 and
    142.3~GHz), spectral index (see text).}
  \label{spec_index}
\end{figure}

Here we examine the distribution of the multi-frequency spectral indices. We define the spectral index
$\alpha$ as $S\sim\nu^{\alpha}$ with $S$ the flux density measured at frequency $\nu$. For each source we
compute mean spectral indices by performing power law fits to averaged spectra.
%
Three-point spectral indices 
have been obtained separately for: (a) the low sub-band over the 4.85, 10.45 and 14.6~GHz, and (b) the high
sub-band at 32, 86.2 and 142.3~GHz). The distributions for the two sub-bands and FSRQ and BL\,Lacs sources
separately, are shown in Fig.~\ref{spec_index}. As it can be seen there:

(i) The low frequency spectral index distribution is shown in the upper panel of Fig.~\ref{spec_index}. The
mean over the whole sample, independently of source class (grey area), is $-0.03$ with a median of
$-0.05$. The FSRQs (black line) show a mean of $0.02$ but their distribution appears rather broadened with a
tail towards more inverted spectra (median of $-0.03$). The BL\,Lacs on the other hand (green line), give a
mean of $-0.08$ (median: $-0.1$). Although their distribution appears shifted slightly towards steeper spectra
as compared to the FSRQs, a K-S test has yielded any significant difference.

(ii) The distribution of the upper sub-band spectral index (32, 86.2 and 142.3~GHz) is shown in the lower
panel of Fig.~\ref{Spec_comb_all}. The mean over the entire sample is still $-0.03$ (median: $-0.13$) though
the overall distribution now appears narrower. We note, however, a few sources contributing values $>0.7$,
i.e. showing, on average, remarkably inverted spectra.  The FSRQs in the sample now show a much narrower
distribution which is interestingly skewed towards negative values with its mean at $-0.23$ (median: $-0.25$).
The BL\,Lacs, however, concentrate around flatter or more inverted spectral indices (mean: $0.35$).
A KS test indicate a significant difference of the two distributions ($P<0.001$).

It must be emphasised that the overall flatness of the broadband radio spectra -- expected for blazars
and discussed above -- is the result of averaging over a ``evolving spectral components'' through the
observing bandpass.

The rather blurred picture seen in the low sub-band spectral index becomes slightly clearer at higher
frequencies. The FSRQs clearly tend to concentrate around negative values contrary to the BL\,Lacs that
concentrate around flat or inverted values. The broader scatter
of the low frequency spectral index is expected since at this regime where the observed emission is the
superposition of slowly evolving past events. At higher frequencies where the evolution is faster this
degeneracy is limited. The divergence of the FSRQs and BL\,Lacs distributions could be understood by assuming
that the latter show turnover frequencies at systematically higher frequencies or that their flares
systematically do not reach the lowest frequencies in contrast to FSRQs. This may indicate different physical
conditions in the two classes. However, this interpretation would also give a natural explanation for the
trend of BL\,Lacs to show lower variability amplitudes (Sect.~\ref{amplitudes}) towards lower
frequencies. 

\section{Radio variability and $\gamma$-ray loudness}\label{fgamma_fermi}


The comparison of the F-GAMMA sample with the early {\it Fermi} AGN source list LBAS
\citep[][]{Abdo_2009ApJ_700_597A}, showed that 29 of our 62 sources were detected already in the first three
months of operation 
confirming the initial source selection. After 11\,months \citep[{\it Fermi} 1LAC catalog
][]{abdo_2010ApJS_188_405A} 54 of the 62 sources (i.e. 87\%) were detected.  As a result, the F-GAMMA program
participated in a several multi-wavelength campaigns initiated by the {\it Fermi} team \citep[e.g. 3C\,454.3,
3C\,279, PKS\,1502+106, Mrk\,421, Mrk\,501, 3C\,66A,
AO\,0235+164;][]{2009ApJ...699..817A,2010Natur.463..919A,2010ApJ...710..810A,2011ApJ...727..129A,2011ApJ...736..131A,2011ApJ...726...43A,2012ApJ...751..159A}
as well as in broadband studies of larger samples \citep[][]{2010ApJ...716...30A,2012A&A...541A.160G}. See
\citet[][]{2010arXiv1007.0348F} for an overview of the early campaigns.

In the following we examine whether the radio variability triggers the source $\gamma$-ray activity and
subsequently their {\it Fermi} detectability.

\subsection{Radio variability amplitude and {\it Fermi} detectability}\label{var_ampl_fermi}
In this section we examine whether radio variability -- expressed by the standard deviation of the flux
density -- is correlated with $\gamma$-ray loudness of the sources. In this context the proxy for the
$\gamma$-ray loudness is the source {\it Fermi} early detectability.

Such a correlation would agree with findings that indicate that $\gamma$-ray flares often occur during
high-flux radio states \citep[e.g.][]{2009ApJ...696L..17K,2011A&A...532A.146L,2014MNRAS.441.1899F}. A
connection between the variability amplitude in the radio (quantified by the intrinsic modulation index) and
the $\gamma$-ray loudness inferred from the source presence in the 1LAC catalog, has been confirmed with high
significance by \citet[][]{2011ApJS..194...29R} using 15~GHz OVRO data. Here, we examine whether such a
connection persists in the F-GAMMA data, and how frequency affects such a connection. Instead of the
modulation index we use the flux density standard deviation.

\begin{figure}
 \centering
   \includegraphics[trim=80pt 10pt 23pt 130pt  ,clip,width=0.3\textwidth,angle=-90]{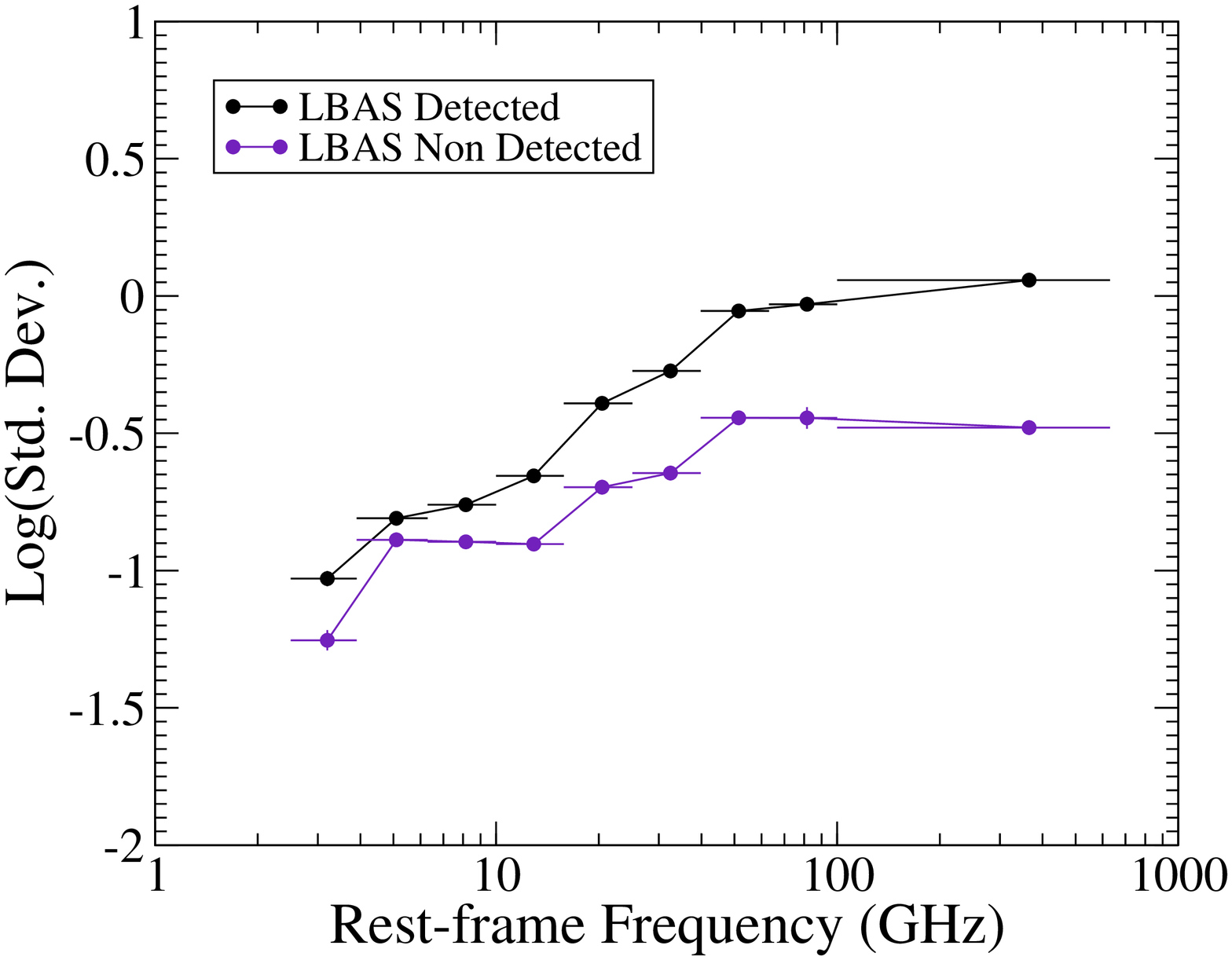}
   \vspace{-.5cm}\\
   \includegraphics[trim=80pt 10pt 20pt 130pt  ,clip,width=0.3\textwidth,angle=-90]{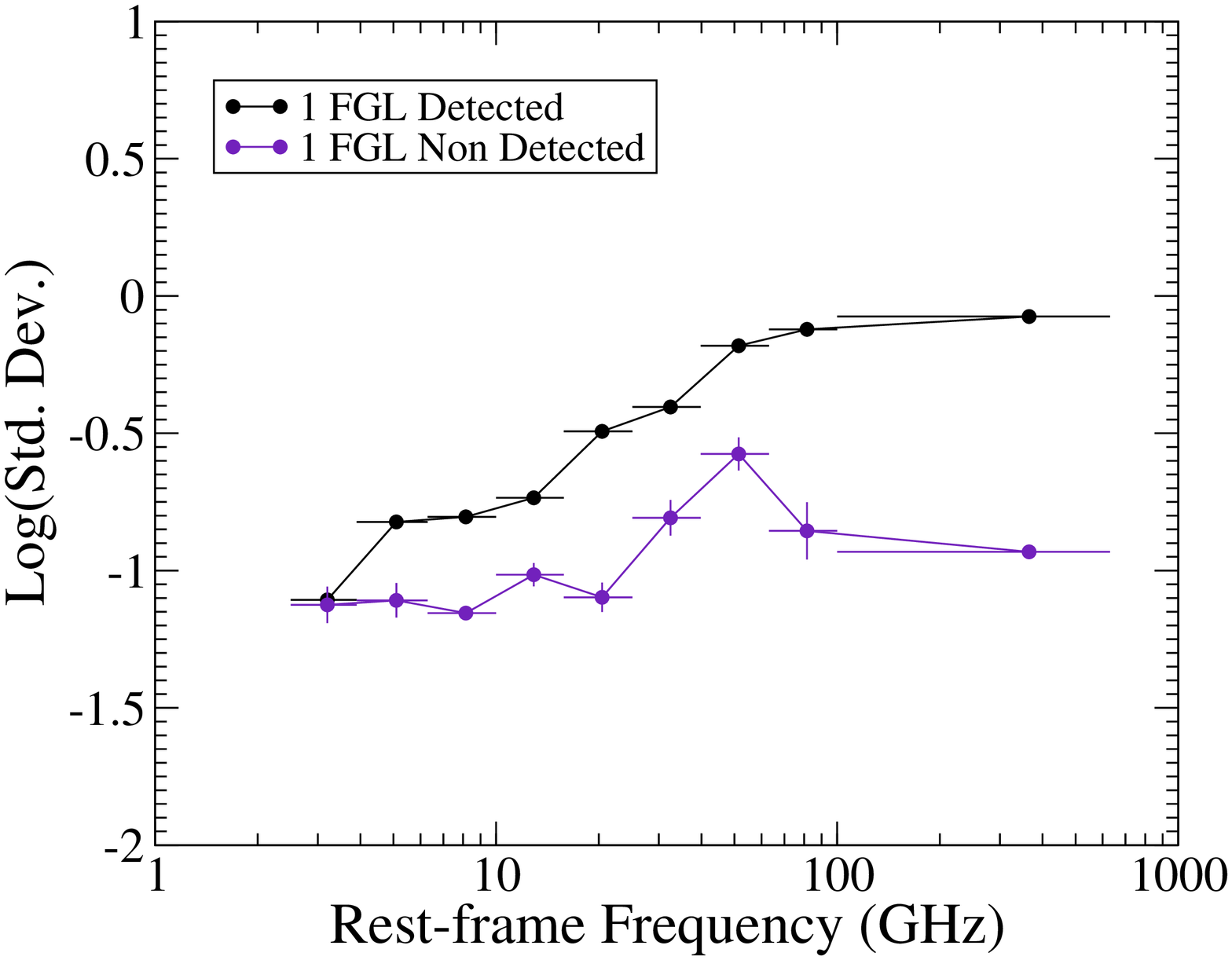}
  \caption{The variability amplitude against rest-frame frequency for the {\it Fermi} LBAS detected/non-detected sources (upper panel) and 1FGL
    detected/non-detected sources (lower panel) of the F-GAMMA sample. {\it y}-axis: logarithmic average of the light curve
    standard deviations }
  \label{gamma_nongamma}
\end{figure}

Figure~\ref{gamma_nongamma} shows the dependence of the logarithmic average of 
the flux density standard deviation on the rest-frame frequency. Sources  
included in one of the first {\it Fermi} catalogs are plotted  separately from 
 those not included. The upper panel refers to the LBAS and the lower one to  
the 1LAC catalog. In the former case (reference to the LBAS), the two curves 
 appear clearly separated. The $\gamma$-ray detected sources display larger 
variability amplitudes confirming our expectations. On average they are more
than a factor of 3 more variable at the highest frequencies where the largest 
separation is seen.  The same conclusion is reached when the 1LAC is used as a 
reference. In this case the statistics is not as good (less F-GAMMA sources are 
not included in the 1LAC), as it is imprinted in our the larger error bars in 
the logarithmic mean. Finally, it is worth noting a clear increase of the 
separation between the two curves
towards higher frequencies. This further supports our findings that the radio/$\gamma$-ray correlation becomes
stronger towards higher frequencies both at the level of average fluxes (see Sect.~\ref{flux_flux_corr}) and
at the level of light curve cross-correlations, when smaller time lags \citep{2014MNRAS.441.1899F}.

\subsection{Brightness temperatures and Doppler factors versus {\it Fermi} detectability} 
We furthermore investigate possible differences between the observed variability time scales, variability
brightness temperatures and Doppler factors of {\it Fermi}-detected and non-detected sources in our sample.

Although no real differences were in the variability time scale, a clear trend of the {\it Fermi}-detected
sources (LBAS and 1LAC) to show variability brightness temperatures larger by a factor of $\sim 2-3$, is
present. The effect is noticeable is all our radio bands.
At 86.2~GHz for example, the LBAS sources show a mean $T_{\mathrm{B,var}}$ of $4.0\times10^{11}$\,K (median:
$9.9\times10^{10}$\,K), whereas the non-LBAS sources exhibit a mean of $1.6\cdot10^{11}$\,K (median:
$9.6\times10^{10}$\,K). Given our previous discussion that {\it Fermi}-detected sources show higher
variability amplitudes (Sect.~\ref{var_ampl_fermi}), such a trend is expected due to the dependence of the
variability brightness temperature to the variability amplitude (Eq.~\ref{tb1}).

We also note a trend of slightly higher variability Doppler factors $\delta_{\mathrm{var,eq}}$ for the {\it
  Fermi}-detected sources across all radio bands. This is in agreement with previous findings of
\citet[][]{2010A&A...512A..24S} reporting larger variability Doppler factors of {\it Fermi}-detected sources
based on Mets{\"a}hovi long term light curves. 

In contrast to \citet[][]{2010A&A...512A..24S}, however, the difference between {\it Fermi}-detected and
non-detected sources in terms of $T_{\mathrm{B,var}}$ and $\delta_{\mathrm{var,eq}}$ discussed above, can not
established with a high statistical significance. It is likely that this is an
effect of (a) limited dataset discussed here, and (b) the method used for the estimation of
$\delta_{\mathrm{var,eq}}$ which relies on average time scales as opposed to time scales of the sharpest,
fastest flares in the light curves.

\section{Radio and $\gamma$-ray flux correlation}\label{flux_flux_corr}
In this section we examine whether an intrinsic correlation between the radio and the $\gamma$-ray fluxes of
sources in our sample exists. That would imply the physical connection between the emission region and
emission processes in the two bands.

Strong correlations were claimed already on the basis of EGRET data
\citep[e.g.][]{1993ApJ...410L..71S,1993MNRAS.260L..21P,1996ApJ...464..600S}, and were re-examined with more
detailed statistical analyses \citep[e.g.][]{1997A&A...320...33M,1998ApJ...496..752C}. A number of
effects however make such correlations uncertain urging for very careful treatment: 
\begin{enumerate}
\item In small samples with limited luminosity dynamic ranges, artificial flux-flux correlations may be
  induced due to the common distance effect.
\item Artificial luminosity-luminosity correlations can emerge when considering objects in flux-limited
  surveys. In such cases most objects are close to the survey sensitivity limit and by applying
  a common redshift to transfer to the luminosity space, artificial correlations appear.
\item The data used to obtain the claimed correlations were not synchronous.
\end{enumerate}

With the large number of {\it Fermi}-detected sources the correlations between radio and $\gamma$ rays have
been revisited over a broad range of radio data
\citep[e.g.][]{2009ApJ...696L..17K,2010MNRAS.407..791G,2011MNRAS.413..852G,2010ApJ...718..587M}.
\cite{2011ApJ...741...30A} used 8~GHz archival data for the largest sample ever used in such studies with 599
sources, as well as a smaller sample of concurrent 15~GHz measurements from the OVRO monitoring program. They
assessed the intrinsic significance of the observed correlations using the data randomisation technique of
\cite{2012ApJ...751..149P}. They confirm a highly significant correlation between radio and $\gamma$-ray
fluxes which becomes more significant when concurrent rather than archival radio data are used.  

The F-GAMMA dataset can provide new insight to the problem owing to some important facts: 
\begin{itemize}
\item The broad frequency range allows us to examine whether the significance and the parameters of the
  correlation show any frequency dependence \citep[see][for a study of this dependence on $\gamma$-ray photon
  energy]{2011ApJ...741...30A}.
\item Our data are perfectly concurrent with measurements of $\gamma$-ray fluxes eliminating biases emerging
  from the non-simultaneity of observations.  
\item The F-GAMMA data provide concurrent information of the radio spectral index, which is an essential input
  for the assessment of the significance of the correlations \citep[][]{2012ApJ...751..149P}.
\end{itemize}

On the other hand, there are certain features of our datasets that require a particularly careful treatment.  First
of all, the sources do not constitute a flux-limited sample. Although this makes them less sensitive to
artificially-induced luminosity-luminosity correlations (Malmquist bias), it also means that statistical tests
usually employed to assess correlation significance can not be benchmarked in a straight-forward way by
sampling the luminosity function \citep[e.g.][]{2008AJ....136.1533B}. As a result, we need a specialised
treatment to estimate how likely it is that a simple calculation of the correlation coefficient will
overestimate the significance of an intrinsic correlation between radio and $\gamma$-ray fluxes due to
common-distance biases, and to calculate the intrinsic correlation significance.

\subsection{The common-distance bias introduced by the limited dynamic range}

As it is shown in \citet[][]{2012ApJ...751..149P}, there is a quantitative criterion that can be applied to
determine the extent to which common-distance bias affects the correlation significance estimated for a
specific dataset using only the value for the correlation coefficient.

The bias is larger for samples with a small luminosity dynamic range, and a large redshift range. Conversely,
samples which have a large luminosity dynamic range compared to their redshift dynamical range are relatively
robust against common-distance biases. This can be immediately understood in the limit where all the sources
are at the same redshift, in which case there is {\em no} common-distance bias.

The quantity summarising the relative extent of the luminosity and redshift dynamic ranges
of a sample is the ratio of the variation coefficient , $c$, of the luminosity and redshift distributions. 
That is defined as the standard deviation in units of the mean. \citet[][]{2012ApJ...751..149P} found that
values of $c_{\mathrm{L}}/c_z$ smaller than $5$ indicate that common-distance biases are important and can
lead to a significant overestimate of the significance of a correlation between fluxes in two bands if only
the correlation coefficient is used, without appropriate Monte-Carlo testing.

\begin{figure*}
  \centering
  \includegraphics[angle=0,width=13cm]{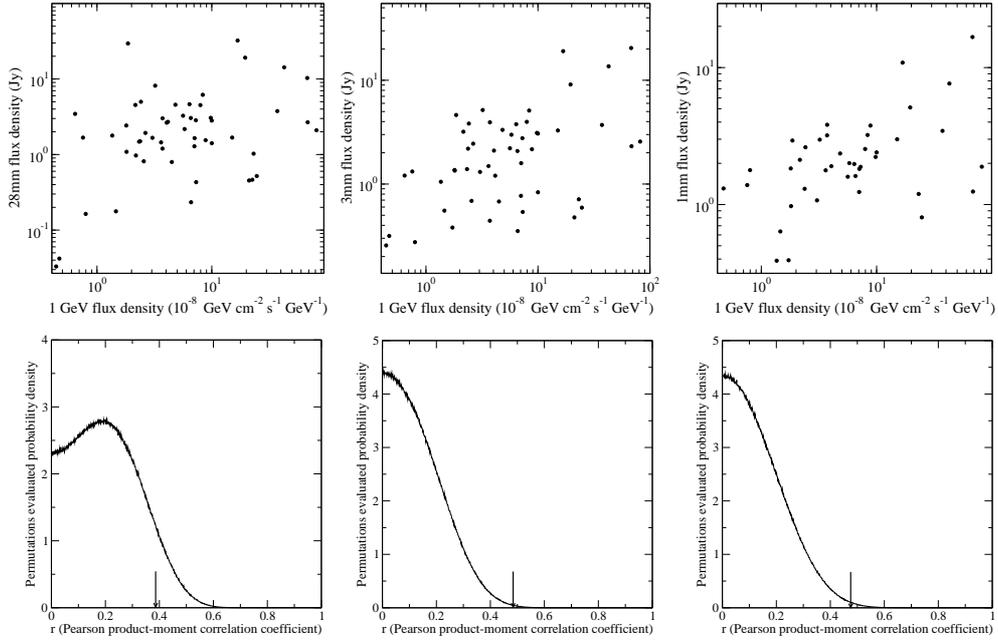}
  \caption{{\bf Top:} Radio flux against {\it Fermi} $\gamma$-ray flux at 10.45~GHz (left), 86.2~GHz (middle)
    and 228.9~GHz (right) for the sources in our sample with known redshifts. {\bf Bottom:} Distribution of
    permutations-evaluated $r-$values (see text) for {\it Fermi} vs. 10.45~GHz (left), 86.2~GHz (middle) and
    228.9~GHz (right) fluxes. Arrows indicate the $r-$values obtained for the actual data.}
  \label{flux-flux}
\end{figure*}

Table~\ref{restable} shows the correlation coefficient for the logarithm of radio and $\gamma$-ray fluxes for
each of our samples (corresponding to a specific radio frequency). As an illustration, the radio and
$\gamma$-ray fluxes are plotted against each other in logarithmic axes for the cases of the 228.9, 86.2 and
10.45~GHz samples in Fig.~\ref{flux-flux}.
\begin{table}
  \caption{Flux-flux correlation analysis: Monte-Carlo results obtained 
    for the different frequencies, where N denotes the number of sources, 
    n the number of redshift bins and $\phi_{\mathrm{high}}$, $\phi_{\mathrm{low}}$ the 
    chance probabilities calculated using the high and low radio spectral 
    index, respectively. See text for details.}
\label{restable}
\centering
\begin{tabular}{rcccccrr}
   \hline
   \hline
\mc{1}{c}{$\nu$} &N &$r$ &$c_{\mathrm{L_\gamma}}/c_{z}$ &$c_{\mathrm{L_r}}/c_{z}$ &n &\mc{1}{c}{$\phi_{\mathrm{high}}$} &\mc{1}{c}{$\phi_{\mathrm{low}}$}\\
\mc{1}{c}{(GHz)} &  &    &                         &                    &  &\mc{1}{c}{(\%)}    &\mc{1}{c}{(\%)} \\
   \hline\\
   228.90 & 41 & 0.47 & 3.95 & 1.57 & 4 & 0.0046 & 0.0049 \\
   142.30 & 51 & 0.51 & 4.38 & 1.76 & 5 & 0.0012 & 0.0011 \\
   86.20  & 52 & 0.48 & 4.33 & 1.76 & 5 & 0.0018 & 0.0018 \\ 
   43.00  & 43 & 0.37 & 4.38 & 1.96 & 4 & 1.9    & 1.6 \\
   32.00  & 47 & 0.19 & 4.37 & 1.95 & 4 & 24.4   & 23.9 \\
   23.02  & 44 & 0.29 & 4.60 & 1.90 & 4 & 6.6    & 6.1  \\
   14.60  & 51 & 0.22 & 4.13 & 1.94 & 5 & 28.1   & 29.3 \\
   10.45  & 53 & 0.38 & 4.10 & 1.93 & 5 & 8.8    & 8.0 \\
   8.35   & 54 & 0.39 & 4.14 & 1.94 & 5 & 5.7    & 5.3 \\
   4.85   & 54 & 0.36 & 4.14 & 1.91 & 5 & 9.8    & 9.3 \\
   2.64   & 53 & 0.31 & 4.18 & 1.92 & 5 & 17.2   & 16.9 \\\\
   \hline
\end{tabular}
\end{table}

As we can see in Table~\ref{restable}, there is a general trend for the correlation coefficient $r$ to be high
at high frequencies ($r \sim 0.5$ for 228.9 to 86.2~GHz), and significantly lower at lower frequencies
($r < 0.4$ at $\nu \leq$~43~GHz). However, these results cannot be taken at face value without appropriate
statistical assessment, because $c_{\mathrm{L}}/c_{z}$ is smaller than 5 for both $\gamma$-ray and
radio frequencies for all of our samples.

\subsection{Treating the limited dynamic range}

To address the peculiarities discussed above \citet[][]{2012ApJ...751..149P} developed a data
randomisation method which is based only on permutations of the observed data. The method preserves the
observed luminosity and flux density dynamic ranges and, provided the sample is large enough, also the observed
luminosity, flux density and redshift distributions. The technique has been designed to perform well even for samples
selected in a subjective fashion, and it has been demonstrated to never overestimate the correlation
significance, while at the same time retaining the power of traditionally employed methods to establish a
correlation when one indeed exists.

In brief, the method has been applied as follows \citep[see][for details]{2012ApJ...751..149P}:

\begin{enumerate}

\item Move to luminosity space using the known redshifts and the relation between monochromatic flux density
  $S_{\mathrm{\nu}}$ and luminosity $L_{\mathrm{\nu}}$.  The simultaneously measured radio spectral indices
  (Sect.~\ref{spec}) allow us to concurrently perform a K-correction and calculate $L_{\mathrm{\nu}}$ at
  rest-frame frequency $\nu_0$ according to:
\begin{equation}
L_{\mathrm{\nu}}(\nu_0)=S_{\mathrm{\nu}}(\nu)\,4\pi d^2(1+z)^{1-\alpha}
\end{equation}
where $d=(c/H_0)\int_0^z dz/\sqrt{\Omega_{\mathrm{\Lambda}}+\Omega_{\mathrm{m}}(1+z)^3}$.  In the case of
$\gamma$-ray observations the actual observed quantity is $F$, the photon flux integrated over energy from
$E_0=1 $\,GeV to $\infty$. This is related to monochromatic energy flux through
$S_{\mathrm{\gamma}} = (\alpha-1)\,F$ where $\alpha$ is the absolute value of the photon spectral index.  The
obtained sets of radio and $\gamma$-ray luminosities fix our luminosity dynamical range.
\item Construct simulated fluxes in radio and $\gamma$-ray by combining each luminosity with one of the
  redshifts. Fluxes outside the original flux range are rejected as e.g. a single very high flux or very low
  flux and a cluster of points of similar fluxes can produce an artificially high correlation index, which
  would not occur given the original flux dynamical range.
\item Pair up the accepted simulated fluxes in all possible combinations excluding the ``true'' flux pairs.
\item Select a large number (${\sim} 10^7$) of $N$ pair combinations, where $N$ 
is equal to the number of the
  original observations. Each set of $N$ pairs is a set of uncorrelated simulated flux observations.
\item Compute the Pearson product-moment correlation coefficient $r$ for each simulated data set.
\item Provided the sample size is large enough, perform steps 2 to 5 in redshift bins, to limit rejection of
  flux values and maintain the luminosity and redshift distributions of the original sample (the sample size
  requirement is to have $\gtrsim$\,10 sources in each bin).
\end{enumerate} 

\subsection{The results of our analysis}

The results of the previous analysis are shown in Table~\ref{restable}. The probability distributions of the
Pearson product-moment correlation coefficient, $r$, for the simulated samples with intrinsically uncorrelated
luminosities are given in Fig.~\ref{flux-flux}. Arrows indicate the $r$-values obtained for the actual data as
given in Table~\ref{restable}.

Radio frequencies at 43~GHz and above have a significant correlation with $\gamma$ rays; better than
$2\sigma$ and, in the case of 142.3 and 86.2~GHz, better than $3\sigma$. Lower frequencies on the other hand
never exceeded $2\sigma$ significance level. These suggest: 
\begin{enumerate}
\item A physical connection between the radio and high-energy emission. In the presence of a low energy
  synchrotron photon field, relativistic electrons and Doppler boosting, such correlation is expected in a
  scenario, where the high energy emission is produced by inverse-Compton (IC)  up-scattering off of low energy
  synchrotron photons.
\item The closer connection of the high radio frequency and IC $\gamma$-ray emitting regions. That is expected
  due to lower intrinsic opacity at mm bands \citep[e.g.][]{2014MNRAS.441.1899F}. 
\end{enumerate} 

The applicability of this result must be appropriately qualified in the light of two specific
concerns. First, because our sample is selected with subjective criteria, the result cannot be generalised to the
blazar population. Instead it is only valid for the specific sources in our sample.

Second, lack of evidence for a significant correlation between low radio ($<43$~GHz) and $\gamma$-ray fluxes
is not equivalent with evidence for lack of a positive correlation. A characteristic counter-example is our
findings at 14.6~GHz. Although no significant correlation was found with our data, a positive correlation has
been established at the same frequency for the {\it Fermi}-detected subset of CGRaBS
\citep[][]{2011ApJ...741...30A,2012ApJ...751..149P} using OVRO 15~GHz data. The reason for this discrepancy is
not the size of the sample, but rather the different makeup of the two samples. There are only 15 sources that
are common to the two samples. Most of the additional sources in our sample are BL\,Lacs, while the OVRO
sample is generally dominated by FSRQs. \citet[][]{2011ApJ...741...30A}, treated BL\,Lacs and FSRQs separately
and found that the correlation between {\it Fermi} and OVRO fluxes for BL\,Lacs was weak. It is only natural
then that the F-GAMMA sample at 14.6~GHz, showed a weak correlation owing to the dominance of BL\,Lacs.

The two last columns of Table~\ref{restable} show the significances calculated using the high
($\phi_{\mathrm{high}}$) and low ($\phi_{\mathrm{low}}$) spectral index discussed in
Sect.~\ref{spec_indices}. The radio spectral index has a mild effect on the calculation of the
significance. We conclude then that the statistical method applied is robust even against small changes of the
radio spectral index.

Finally, we have tested whether the time duration of integration affects the strength and the significance of
the correlation between radio and $\gamma-$ray fluxes. For the 28 sources common in our 86.2~GHz sample, the
LBAS and 1FGL, we have calculated the correlation coefficient $r$ and the significance of the correlation
between 86.2~GHz and 1~GeV flux densities averaged over three months and one year. Those are time spans
relevant for LBAS and 1FGL, respectively. In the first case (LBAS) we found that $r=0.5$ and p-value of
$5.9\times10^{-3}$. In the second case (1LAC), we found that
$r=0.44$ with a significance of 4.3~\%. We conclude that that the correlation weakens when the averaging is
extended over significant longer time periods ($\sim$ a year).  Since the typical flaring event duration in
radio is a couple of months, and assuming that over short $\gamma$-ray integrations it is typically the
flaring sources that are detected, this effect may be and indication that there is a common origin between GeV
flares and flares at high radio frequencies.

\section{Summary and conclusions} 
\label{sect:Summary}

We have presented the {\it Fermi} dedicated blazar radio multi-frequency monitoring program, F-GAMMA. The
F-GAMMA program conducted a monthly monitoring of the radio variability and spectral evolution of $\sim60$
$\gamma$-ray blazars at 12 frequencies between 2.6 and 345~GHz. The observations were carried with the
Effelsberg 100-m, IRAM 30-m and APEX 12-m telescopes including polarisation at several bands. The initial
sample presented here has been selected from the most prominent, frequently active, and bright blazars at
$\delta \ge -30^\circ$. The conclusions of the first 2.5 years analysis can be summarised as follows:

\begin{itemize} 
\item Our analysis showed that almost all sources are variable across all frequency bands. On the basis of a
  maximum likelihood analysis that accounts for possible biases, we have demonstrated that the variability
  amplitude increases with increasing frequency up to rest-frame frequencies of $\sim60$ -- 80~GHz; Above that
  the variability decreases or remains constant. The variability of individual sources, however, can rise
  continuously across or peak within our band.
  These findings agree with predictions of shock-in-jet models where the maximum amplitude of flux variations
  is expected to follow the standard growth, plateau and decay phase. 

\item At lower frequencies the FSRQs in our sample show larger variability amplitudes than BL\,Lacs -- in
  terms of flux density variance -- in contrast to previous findings that used variance in units of mean flux
  density. The discrepancy arises form the frequency dependence of the flux density which for BL\,Lacs is on
  average significantly lower. This leads to apparently higher variability amplitudes for BL\,Lacs when the
  mean flux density is used for the normalisation of the variance.

\item The variability time scales range from 80 and 500 days depending on source and frequency. A clear trend
  of faster variability towards higher frequencies is observed. As an example mean values of 348, 294 and 273 days
  at 2.64, 14.6 and 86.2~GHz, respectively have been measured.

\item 
  The calculated $T_{\mathrm{b,var}}$ values depend on frequency. They typically range from $10^{9}$ to
  $10^{14}$~K. A systematic trend of decreasing $T_{\mathrm{b,var}}$ (by two orders of magnitudes) and
  $\delta_{\mathrm{var,eq}}$ (by more than a factor of 4) towards higher frequencies is observed with mean
  values for $\delta_{\mathrm{var,eq}}$ of e.g. 8.8, 4.8 and 2.3 at 2.64, 14.6 and 86.2~GHz, respectively.

\item The combination of EB and PV datasets has resulted a large data base of monthly sampled broadband
  spectra. Their time coherence is kept at 10 days and below. 
  Typically an evolving synchrotron self-absorbed component over a low-frequency steep-spectrum component
  (quiescent jet) is observed. Often the spectra follow the standard three-stage evolutionary path of shock. A
  physically different mechanism appears likely for several sources that display nearly and achromatic
  variability.
  The spectral evolution can also explain naturally the general flatness of the spectra (mean spectral index
  $-0.03$ at low as and high frequencies). The spectral flatness is resulting naturally from averaging
  over different spectral components and their evolution across the spectrum over the 2.5-year-long period.

\item We find significant differences between the FSRQs and BL\,Lacs in our sample. The BL\,Lacs show
  systematically lower variability amplitudes at lower frequencies. The difference vanishes at higher
  frequencies. Although the variability time scales appear similar, the variability brightness temperatures
  and Doppler factors are also lower, at lower frequencies, for BL\,Lacs. The difference is again decreasing
  towards higher frequencies.  This behaviour can be understood in the light of our spectral findings. For
  BL\,Lacs the high frequency spectral indices are flatter or more inverted. This implies that flares
  appearing in BL\,Lacs show higher turn-over frequencies and systematically not reaching the lowest bands.
  Subsequently, they invoke lower variability amplitudes, lower $T_{\mathrm{b,var}}$ and lower
  $\delta_{\mathrm{var,eq}}$ values at these bands.

\item We have searched for possible correlations of radio characteristics with $\gamma$-ray loudness. As proxy
  for the latter we have used the presence of the sources the LBAS and 1LAC catalogs. We find that the {\it
    Fermi}-detected sources show larger variability amplitudes than non-detected ones. The clear increase of
  the separation between flux standard deviation averages with increasing frequency is supporting an arguably
  tighter correlation between $\gamma$ rays and higher radio bands. We also find a trend of higher
  $T_\mathrm{b,var}$ and $\delta_{\mathrm{var,eq}}$ in {\it Fermi}-detected sources.

\item We have searched for correlations between F-GAMMA flux densities and concurrent {\it Fermi}/LAT 1~GeV
  fluxes for frequencies even up to 228.9~GHz. After a careful treatment of the limited dynamic ranges of our
  sample, we find that: flux densities at $\nu\geq$43~GHz correlate with 1~GeV fluxes at a significance level
  of better than $2\sigma$; at 142.3 and 86.2 GHz the significance is better than $3\sigma$. 
  This implies that the $\gamma$-ray emission is produced very close to the mm-band emission region. This view
  is also supported by the fact that at 86.2~GHz flux densities averaged over a few months (comparable to the
  duration of a single flare), correlate at higher significance than flux densities averaged over longer time
  scales (a year as in the 1LAC catalog).
\end{itemize}

A five-year data analysis based on the revised F-GAMMA sample will be presented in a subsequent
publication. Nestoras et al. (submitted) discuss the first five years of PV data. \cite{2012JPhCS.372a2007A}
studied the variability of the radio broadband spectra. They proposed that the variability patterns can be
classified in merely two categories. Those showing intense spectral evolution and those that vary
achromatically. They show that the former can be easily reproduced with simply the superposition of a
steep-spectrum steady state component and a high frequency one that evolves in time and frequency. Concerning
the $\gamma$-ray emission site, \cite{2014MNRAS.441.1899F} conducted a cross-correlation analysis between
mm-radio and $\gamma$ rays. They found that the $\gamma$-ray emission for 3C\,454.3 originates at a distance
of at least 0.8--1.6~pc from the supermassive black hole. \cite{2016A&A...586A..60K} examined the structural
evolution of PKS\,1502+106 during a $\gamma$-ray outburst with mm-VLBI. Using F-GAMMA data they estimated that
the $\gamma$-ray emission site must be no farther than ${\sim} 6$~pc from the jet base. Later
\cite{2016A&A...590A..48K} localised the emission site at $1.9\pm1.1$~pc from the jet
base. \cite{2014arXiv1401.2072M} looked at the multi-frequency linear and circular polarisation data to
interpret rotations of the polarisation angle in terms of an opacity evolution 
effect. Liodakis et
al. (submitted) study the variability Doppler factors for EB and PV data and compares them with the
predictions of blazar population models.

\begin{acknowledgements}
  The authors would like to thank the anonymous referee for the very careful reading of the manuscript, and
  the constructive and insightful comments. They have noticeably improved the quality of the paper. We would
  also like to thank the MPIfR internal referees Drs J. Hodgson and C. Casadio for the careful reading of the
  manuscript. This research is based on observations with the 100-m telescope of the MPIfR
  (Max-Planck-Institut f\"ur Radioastronomie) at Effelsberg. It has also made use of observations with the
  IRAM 30-m telescope. IRAM is supported by INSU/CNRS (France), MPG (Germany) and IGN (Spain). IN, VK and IM
  were funded by the International Max Planck Research School (IMPRS) for Astronomy and Astrophysics at the
  Universities of Bonn and Cologne. The OVRO 40-m program is supported in part by NASA grants NNX08AW31G and
  NNG06GG1G and NSF grant AST-0808050. We would finally like to acknowledge partial support from the EU FP7
  Grant PIRSES- GA-2012-316788.
\end{acknowledgements}

\bibliographystyle{aa}
\bibliography{references.bib}

\end{document}